\documentclass[twocolumn,tighten,times]{aastex7}
\usepackage{color}
\usepackage{graphicx}
\usepackage{xcolor}
\usepackage{natbib}

\newcommand{\target}{{PKS\,0023$-$26}}

\newcommand{\kms}{$\,$km$\,$s$^{-1}$}

\newcommand{\ergs}{$\,$erg$\,$s$^{-1}$}

\newcommand{\msunyr}{{$M_\odot$ yr$^{-1}$}}

\newcommand{\pks}{{PKS\,0023$-$26}}
\newcommand{\coOne}{{CO(1-0)}}
\newcommand{\coTwo}{{CO(2-1)}}
\newcommand{\coThree}{{CO(3-2)}}

\newcommand{\rdt}{$R_{\rm 32}$}

\def\emph#1{{\sl #1}}
\newcommand{\ltsima} {$\; \buildrel < \over \sim \;$}
\newcommand{\gtsima} {$\; \buildrel > \over \sim \;$}
\newcommand{\lta} {\lower.5ex\hbox{\ltsima}}
\newcommand{\gta} {\lower.5ex\hbox{\gtsima}}

\newcommand\chandra{{\sl Chandra}}

\newcommand\sherpa{Sherpa}

\newcommand\xmm{XMM-{\it Newton}}

\usepackage{txfonts}
\begin{document} 

\title{The changing impact of radio jets as they evolve. \\
   Part II: Chandra study of the X-ray emission of \pks\ and comparison with recent ALMA results.}
\shorttitle{Chandra study of \pks}
\shortauthors{Siemiginowska et al.}

    \author[0000-0002-0905-7375]{A. Siemiginowska}
    \affiliation{Center for Astrophysics $|$ Harvard \& Smithsonian, Cambridge MA 02138 USA}
    \email{asiemiginowska@cfa.harvard.edu}
    \correspondingauthor{A. Siemiginowska}
    \email{asiemiginowska@cfa.harvard.edu}

    \author[0000-0002-9482-6844]{R. Morganti}
    \affiliation{ASTRON, the Netherlands Institute for Radio Astronomy, Oude Hoogeveensedijk 4, 7991 PD, Dwingeloo, The Netherlands} 
    \affiliation{Kapteyn Astronomical Institute, University of Groningen, Postbus 800,
9700 AV Groningen, The Netherlands}
    \email{morganti@astron.nl}

    \author[0000-0002-3554-3318]{G. Fabbiano}
    \affiliation{Center for Astrophysics $|$ Harvard \& Smithsonian, Cambridge MA 02138 USA}
    \email{gfabbianoa@cfa.harvard.edu}

    \author[0000-0002-5671-6900]{E. O'Sullivan}
    \affiliation{Center for Astrophysics $|$ Harvard \& Smithsonian, Cambridge MA 02138 USA}
    \email{eosullivan@cfa.harvard.edu}

    \author[0000-0002-0616-6971]{T.Oosterloo}
        \affiliation{ASTRON, the Netherlands Institute for Radio Astronomy, Oude Hoogeveensedijk 4, 7991 PD, Dwingeloo, The Netherlands} 
        \affiliation{Kapteyn Astronomical Institute, University of Groningen, Postbus 800,
9700 AV Groningen, The Netherlands}
    \email{}
    
    \author[0000-0002-2951-3278]{C.Tadhunter}
    \affiliation{Department of Physics and Astronomy, University of Sheffield, Sheffield, S7 3RH, UK}
    \email{}
    
    \author[0000-0003-4428-7835]{D. Burke}
    \affiliation{Center for Astrophysics $|$ Harvard \& Smithsonian, Cambridge MA 02138 USA}
    \email{dburke@cfa.harvard.edu}

    
   \title{Chandra X-ray Observatory study of the X-ray emission of \pks\ \\
   and comparison with recent ALMA results.}

\begin{abstract}

We present a deep high-resolution \chandra\ X-ray Observatory image data of a
powerful compact radio source  \pks\ associated with a quasar at redshift 0.322. The earlier studies of the optical environment suggested that the source could be located in a galaxy cluster or a group.
However, we report a non-detection of hot gas on large scales (out to $\sim 60$\,kpc radius) and place an upper limit on the X-ray luminosity of $<3\times10^{42}$\,erg\,s$^{-1}$, consistent only with the presence of a poor, low-temperature ($\rm kT < 0.5$\,keV) galaxy group.
X-ray spectral analysis of the central circular region, $r<7$\,kpc shows, in addition to the mildly absorbed AGN, a thermal emission component with a temperature of $\rm kT=0.9^{+0.19}_{-0.37}$\,keV. We discuss the origin of this hot component as a result of interaction between the evolving radio source and the interstellar medium.
Our high angular resolution X-ray image traces the distribution of hot gas which is closely aligned with and extends beyond the radio source, and also in the direction perpendicular to the radio source axis. 
The X-rays are enhanced at the northern radio lobe and the location of the peak of the \coThree/\coTwo\, line emission, suggesting that the interactions between the jet and cold medium result in the X-ray radiation which 
excites CO. The shock driven by the jet into the ISM is supersonic with the Mach number of $\mathcal{M} \sim 1.75-2$, creating the cocoon of hot X-rays surrounding the radio source. This result agrees with observations of shocks in other radio galaxies pointing to a prevalent impact of jets on ISM. 

\end{abstract}
   \keywords{active galaxies (17); radio galaxies (1343); radio jets (1347).}


\section{Introduction}
\label{sec:introduction}

It is well established that relativistic jets emerging from active galactic nuclei (AGN) are powered by accretion onto supermassive black holes. 
However, the trigger mechanism of a jet and its connection to the feedback process are not well understood. 
Recent studies of morphologies \citep[e.g., see][]{Ramos2011,Ramos13,Pierce2022} show strong evidence that the most luminous radio AGN are triggered by galaxy mergers and interactions.
The impact of jets on the interstellar environment may limit the fuel supply and thus regulate black hole activity and its growth \cite[e.g.][]{DiMatteo2005, Mukherjee16, Hopkins2016, Cielo2018}.

Radio jets span a broad range of scales, enabling them to influence the surrounding medium from parsec (pc) to hundreds of kiloparsecs (kpc). Their large-scale impact is reflected in the morphology of X-ray clusters via large cavities, filaments or shocks \citep[e.g. see reviews by][]{Fabian2012,McNamara2012}.    
Jets can also drive powerful molecular \citep[and refs therein]{Morganti2015,Morganti2018,Ruffa2024}
and warm, emission-line outflows \cite[e.g.,][]{Tadhunter1991,Emonts2005,Nesvadba2006,Mahony2016,Jarvis2019},
that empty the central regions of galaxies and limit the black hole fuel supply. ALMA observations of nearby X-ray clusters show a rich molecular gas morphology dominated by filamentary or disk-like structures,
suggesting a dependence between the distribution of the cold gas and the jet activity \citep{Russell2019}. 
Most of these clusters have well-developed large-scale radio jets and lobes indicating multiple and ongoing phases of radio activity. Also in many clusters the X-ray morphology implies that there have been multiple phases of jet activity with timescales as low as $\sim$Myr, much shorter than the merger timescales of $\sim$0.2-1\,Gyr derived from cosmological simulations 
\citep{lotz2011,Sabater2019}.

X-ray clusters undisturbed by multiple large scale radio outbursts, such as those associated with known compact radio sources, are rare. For example, a sample of 46 Compact Steep Spectrum (CSS) radio sources which are fully embedded in their host galaxy ISM
\cite[see,][for a review of the CSS sources]{O'Dea1998,O'Dea2021}
have been observed by \chandra\ 
\citep{Siemiginowska2008,Kunert-Bajraszewska2014,Sobolewska2019}
and only two CSS sources have been found in bright X-ray clusters
(3C186,\,$z$=1.06, \citealp{Siemiginowska2005,Siemiginowska2010}; 1321+045,\,$z$=0.312, \citealp{Kunert2013,O'Sullivan2021}). In both cases the small-scale ($<10$ kpc) CSS radio source is located at the center of a luminous ($L_X > 10^{44}$\ergs) cool-core cluster with no detectable signatures of past radio outbursts.
Recent studies of two clusters hosting young ($<10^3$ yrs) radio galaxies by \cite{Ubertosi2023} reveal that the 'pre-feedback' clusters exhibit lower central entropy and cooling times compared to clusters with evolved radio sources.

Observations of warm ionized gas or the cold molecular component show that young radio jets can drive fast outflows and increase the turbulence of the surrounding gas \citep[e.g.,][]{Holt2003,Holt2008,Guillard12,Tadhunter2014,Harrison18,Santoro20,Ruffa24,Cresci23,Costa24}.
The powerful jets can shock the interstellar medium (ISM) resulting in aligned X-ray radiation, as observed in nearby radio galaxies (e.g. in 3C171, \citealt{Hardcastle2010a}: 3C305, \citealt{Massaro2009,Hardcastle2012}; PKS B2152-69, \citealt{Worrall2012},
3C303.1 \citealt{O'Dea2006,Massaro2010}, 3C237 \citealt{Massaro2015}, and CSS sources in \citealt{O'Dea2017}).

Additionally theoretical models predict that small-scale radio source lobes can emit relatively strong X-rays via inverse Compton scattering on the hot relativistic plasma \citep{Stawarz2008b,Krol2024}.
\chandra\ observations have shown that jets interact with the host galaxy ISM, causing either shock-ionization (e.g. \citealt[][in NGC 4151]{Wang2011}; \citealt[][in MRK 573]{Paggi2012})
or compression, resulting in enhanced, cooler X-ray emission \cite[][in NGC6117]{Fabbiano2022}.
Embedded jets in ISM - rich galaxy disks may also cause lateral outflows and hot X-ray halos 
\citep{Mukherjee16,Mukherjee2018}
that have been detected in nearby AGNs with Chandra and HST (e.g., in ESO 428-G014 \citealt{Fabbiano2018};
in IC 5063, \citealt{Travascio2021,Maksym2020}; and in NGC 6117, \citealt{Fabbiano2022}, see also review by \citealt{Fabbiano22a}).

In order to study the impact of a young radio source on the ISM of its host galaxy, and also connect to the larger scale environment we obtained a deep \chandra\ observation of
\pks\, a double-lobe CSS radio source 
at redshift $z = 0.32188(4)$ \citep{Santoro20}
with size of 
$\sim$3\,kpc \citep[peak-to-peak lobes; a full extent up to $\sim 4.7$ kpc;][]{Tzioumis02} and
a radio core detected in recent ALMA observations
\citep{Morganti2021}.
The powerful radio source ($\log P_{\rm 5\, GHz}/{\rm W\, Hz^{-1}} = 27.43$) is
hosted by an early-type galaxy \citep{Ramos2011} and it
is classified as Type 2 quasar with a bolometric luminosity in the range of $2.5 - 4 \times 10^{45}$\ergs\  \citep{Holt2008,Santoro20}.

There are several early-type galaxies located $<30\arcsec$\,  away from \pks, including three galaxies at similar redshift located in projection at 5.8\arcsec (27\,kpc), 6.9\arcsec (32.5\,kpc), 25.8\arcsec (116\,kpc) distance
suggesting that \pks\ could be located at the heart of a rich cluster of galaxies \citep{Tadhunter2011,Ramos13}.
On the other hand, based on short \xmm\ observations, the total X-ray luminosity of \pks\ and its surroundings ($\log L_{\rm 2-10\,keV}/{\rm erg\, s^{-1}} = 43.27$; \citealt{Mingo14}) is lower than expected for a rich cluster of galaxies, and is more consistent with that of a galaxy group \citep{Eckmiller11}.
The questions about the environment of \pks\, and the impact of the radio source on the ISM has motivated the new \chandra\, X-ray observations presented in this paper. 

Recent ALMA studies of this quasar (\citealt{Morganti2021, Oosterloo2025}, Paper I) provide a comprehensive characterization of the cold molecular gas. Together with the \chandra\ results, we can now compile a multi-phase picture of the environment of \pks.

We obtained a deep \chandra\ observation of \pks\ and aimed to detect the X-ray gas associated with group or cluster, evaluate X-ray morphology of the hot gas on sub-arcsec scales and study the distribution of the hot gas in relation to the molecular gas. Our goal was also to study the spectral properties of the central AGN and the impact of a young radio jet on the ISM on smaller radial scales.
We did not detect an X-ray luminous cluster, but we found hot gas surrounding the radio source indicating interactions between evolving jet and the ISM.
We present the \chandra\ observations and data analysis in Section~\ref{sec:chandra}, describe the results in Section~\ref{sec:X-ray} and 
in Section~\ref{sec:discussion} we discuss our results in the light of the radio and molecular ISM properties.

We adopted the flat Universe cosmology with $H_{\circ} = 70$ \kms\ Mpc$^{-1}$, $\Omega_\Lambda = 0.7$, $\Omega_{\rm M} = 0.3$ \citep{Hinshaw2013}; 1\arcsec corresponds to 4.716\,kpc at $z=0.322$ redshift of \pks.

\section{Chandra observations and Data Analysis}
\label{sec:chandra}

 \begin{figure}
 \vskip 0.2in
   \centering
   \includegraphics[angle=0,height=5.5cm]{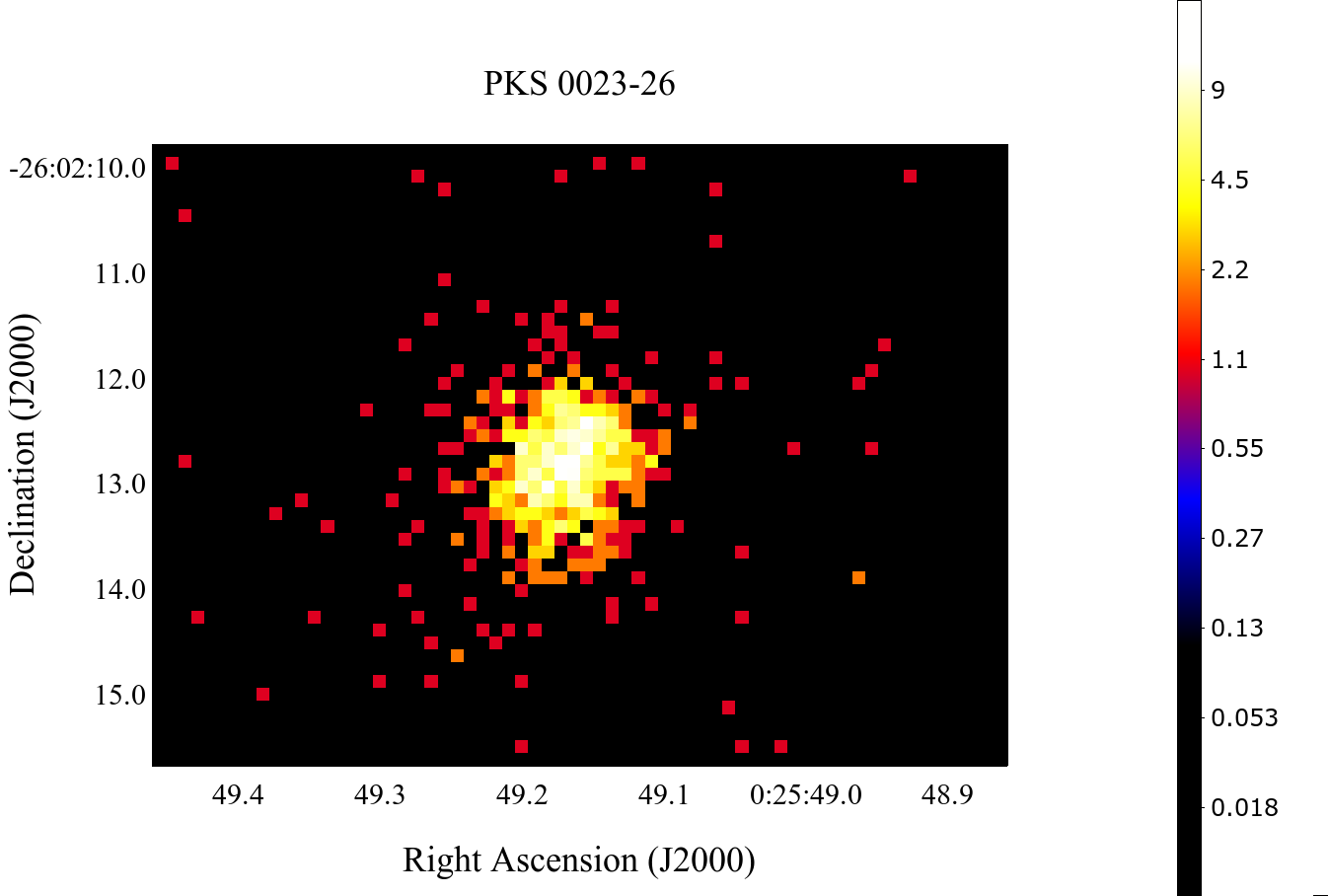}
   \caption{\chandra\, ACIS-S image of \pks\, in the 0.3-7\,keV energy range. The X-ray image is binned to 1/4th ACIS pixel size (0.\arcsec123). The scale is log and the color indicates counts in a pixel as shown on the bar to the right.}
  \label{fig:x-ray-image-broad}
    \end{figure}

\begin{table}
{
\caption{\label{tab:obsids}\chandra\ Observations}
\begin{center}
\hfill
\vskip -0.1 in
\begin{tabular}{cccrcrrl}\hline\hline
 Date & obsid$^a$  & Exposure$^b$ &  net counts$^c$  \\
 & &  (ks) &  \\
\hline
2021-09-13 & 24302 & 27.5 & $125.3\pm11.6$\\
2021-09-19 & 24301 & 17.6 & $98.8\pm10.2$\\
2021-09-20 & 23839 & 24.5 & $128.1\pm11.6$\\
2021-09-21 & 26134 & 28.7 & $95.0\pm10.1$ \\
2021-09-23 & 24300 & 32.6 &  $130.5\pm11.7$\\
2021-09-24 & 24303 & 29.0 & $129.4\pm11.7$ \\
\hline
\hline
\end{tabular}
\end{center}
}
{\footnotesize Notes: 
$^a$ \chandra\ obsid;
$^b$ effective exposure time for each obsid;
$^c$ number of background subtracted counts in the 0.5-7~keV energy range in a r = 1.5\arcsec\, circular region centred on the source.}
\end{table}

The \chandra\ observation of \pks\ was performed in September 2021 with the Advanced CCD Image Spectrometer (ACIS-S) and the readout of the full CCD in the VFAINT mode.  The source was placed on the S3 chip at the nominal aim-point for the Cycle 22 observations. The total exposure time of $\sim$160\,ksec was split into six separate pointings with exposure times from 17.6\,ksec to 32.6\, ksec. 
We list information about each pointing in Table~\ref{tab:obsids} including the obsid, effective exposure time and measured source counts. 

We used CIAO v.4.16 \citep{Fruscione2006} and the calibration CALDB v.11.2 for the X-ray data analysis. 

\subsection{Image Analysis}
\label{sec:image-analysis}

The deep Chandra observation of \pks\ was split into six individual pointings, carried out over the course of 10 days (see Figure~\ref{fig:x-images} in the Appendix showing individual observations). The source flux did not vary during that time within the 90\% measurement uncertainties (see Table~\ref{tab:fit-obsids}). 
We applied the astrometric correction with {\tt wcs\_match, \tt wcs\_update} and {\tt reproject\_events} tools to each data set before merging the six event files with {\tt dmmerge}. 
The merging of these observations resulted in a deep X-ray image with a total exposure time of 160 ksec, shown in Figure ~\ref{fig:x-ray-image-broad}, providing the highest S/N to search for possible diffuse emission surrounding \pks.

\chandra\ X-ray images have the highest angular resolution attainable to date and, depending on the observation setting, they can achieve a sub-pixel resolution, i.e. $< 0.\arcsec492$ for the ACIS-S detector. In our analysis we used {\tt chandra\_repro} and applied the EDSER repositioning algorithm \citep[{\tt {pix\_adj=EDSER}},][]{Li2004} which improves the photon localizations and results in a 40\%-70\% narrower width of the default point spread function (PSF). This allows for the analysis of on-axis images at the bin size of 0.\arcsec0615 \citep[see e.g.][]{Wang2011}. However, the angular image resolution depends on the shape of the PSF\footnote{https://cxc.cfa.harvard.edu/ciao/PSFs/psf\_central.html} at a given off-axis location of the source. Thus we perform the PSF simulations (see Sec.~\ref{sec:psf-sims} below) to evaluate
the angular resolution of the X-ray images we discuss in the following sections.

 \begin{figure*}
 \vskip 0.2in
   \centering
   \includegraphics[angle=0,height=5.5cm]{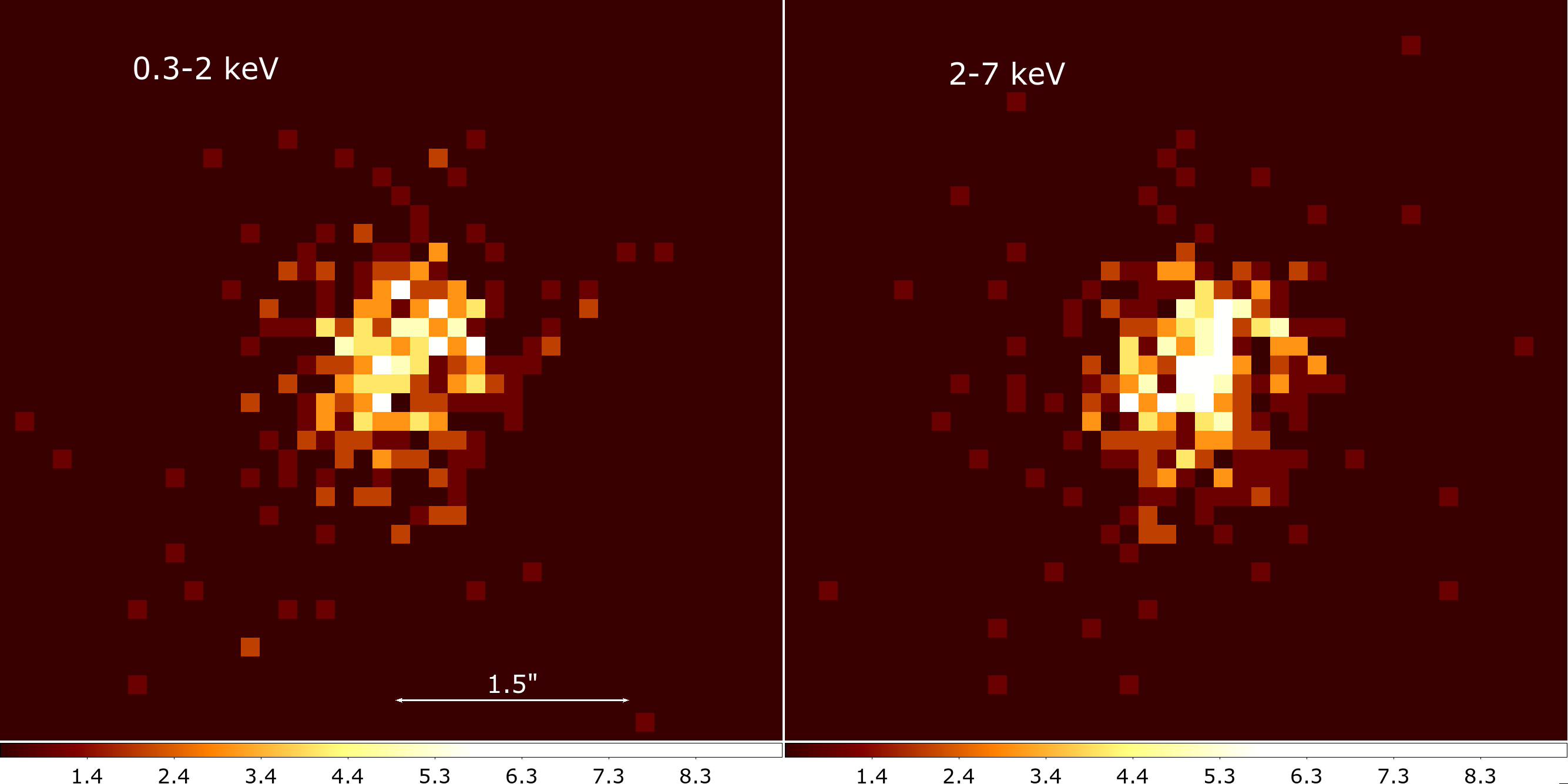}
\caption{\chandra\, ACIS-S image of \pks\, in two bands: soft energy bend, 0.3-2\,keV (left) and hard energy band 2-7\,keV (right). The X-ray image is binned to 1/4th ACIS pixel size (0.\arcsec123). The arrow shows 1.\arcsec5 scale and the size of each image is $5.\arcsec9\times5.\arcsec9$. The color indicates counts in a pixel as shown on the color bars. }
  \label{fig:x-ray-soft-hard}
    \end{figure*}

Figure~\ref{fig:x-ray-soft-hard} shows two X-ray images of \pks\ with the counts accumulated in the soft (0.3-2\, keV) and hard (2-7 keV) energy range respectively. The images are binned to 0.\arcsec123 pixel size. The hard image has a more defined core emission and some scatter around it. The soft image shows a more uniform distribution of counts with no specific core visible. Such difference in morphology is expected for an absorbed AGN -  the core visible in the hard band and absorbed in the soft band with the soft X-rays originating outside the primary core region. We will perform the spectral analysis to quantify the parameters of the X-ray radiation (see Sec.~\ref{sec:xray-spectrum}).

\subsubsection{PSF Simulations}
\label{sec:psf-sims}

   \begin{figure}
   \centering
    \includegraphics[width=0.95\columnwidth{}]{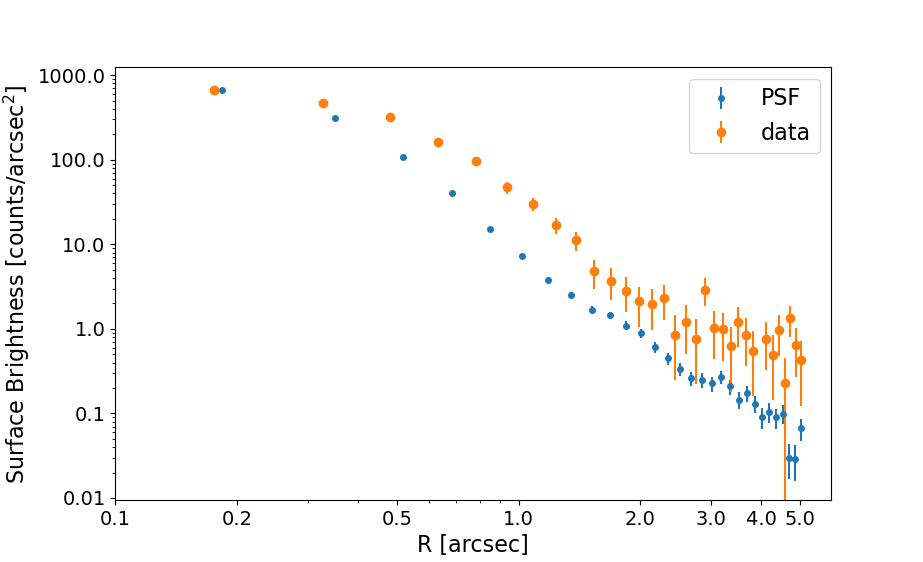}
\caption{The surface brightness profiles of the \pks\ X-ray image (0.5-7 keV) and the \chandra\ PSF. The \pks\ profile is marked in orange and the \chandra\ PSF in blue points.}
    \label{fig:profile}
    \end{figure}

The peaked PSF is energy dependent and contains about 50\% counts within a compact radius of about $<0.\arcsec25$ at soft energies $<1.5$\,keV (on-axis)\footnote{Chandra Observatory Guide: {\url{https://cxc.harvard.edu/proposer/POG/html}}}. 
However, the shape of the \chandra\ PSF varies widely across the field of view and it cannot be described analytically, therefore
the high resolution image analysis requires the PSF simulation with CIAO tools, CHART and MARX. We input the best fit spectrum, the source coordinates for each observation and the aspect solution files to CHART and generated 50 random PSF rays which were projected onto ACIS-S detector using MARX (with the corresponding aspect dither and specific coordinates for each observation RA\_NOM, DEC\_NOM, ROLL\_NOM) and merged into one PSF for specific {\tt obsid}. We added an intrinsic broadening of the PSF using the BLURR parameter in MARX set to 0.\arcsec07 \citep[see][ for a discussion on the Chandra PSF]{Ma2023}. We generated the final PSF by merging the PSF simulated for each {\tt obsid}. 

We extracted the surface brightness profiles of the X-ray emission and the PSF assuming annular regions centered on the peak emission.
The observed surface brightness profile of \pks\ is broader than the PSF at the scales larger than $>0.\arcsec32$, thus not consistent with the point source emission (see Figure~\ref{fig:profile}). Note that this is also a rough estimate of the angular resolution of the images we discuss in Section~\ref{sec:X-ray-image}.


\subsection{Spectral Analysis}

We used {\tt specextract} in CIAO to obtain the X-ray spectrum of \pks\ in each individual observation assuming a circular region with $r=1.5\arcsec$ ($\sim$7\,kpc, PSF fraction of 94\%) radius centered on the source. The background spectra were extracted from the annulus centered on the source with the inner and outer radii of 2.8\arcsec\ and 5.8\arcsec\ respectively. The associated response files and effective area files were generated for each observation. The six spectra and the responses were combined into one spectrum using the {\tt combine\_spectra} script resulting in a total of 692.3$\pm$26.5 net counts. We model this spectrum in Sherpa \citep{Freeman2001,Sherpa2024} using the fit statistic {\tt cstat} based on the Poisson likelihood (see Sherpa documentation for details).

The spectral extraction region, $r=1.5\arcsec$, includes the emission of the AGN nucleus, an intrinsic absorption associated with the nucleus/core, radiation scattered off the molecular medium in the nucleus/core and addition of diffuse emission on larger scales, but still within $<1.5\arcsec$ ($<7$\, kpc) distance from the nucleus. An AGN X-ray spectrum associated with the accretion flow and the hot corona is characterized by a power law. This radiation could be absorbed by a large amount of gas, and it can also be scattered off the molecular medium resulting in the narrow iron line, Fe-K$_{\alpha}$, emitted at E=6.4\,keV \citep[see review by][]{Hickox2018}. 
We applied 
several models with increasing complexity and number of parameters to the combined X-ray spectrum of \pks. 
The models defined in \sherpa\ were used to build model expressions: (1)  A simple power law model, {\tt powlaw1d},  defined on energy scale E, $M(E) = \rm Norm*\rm E^{-\Gamma}$ with the normalization at 1\, keV in units of $[\rm photons\,cm^{-2}\,s^{-1}]$ and a photon index $\Gamma$; (2) The Galactic absorption model, {\tt phabs}, a multiplicative expression
$\exp[-\rm N_H\sigma_E]$ with the equivalent hydrogen column $\rm N_H$, and the atomic cross sections $\sigma_E$ from \citep{Anders1989};  
(3) The intrinsic absorption, {\tt zphabs}, at the redshift of the source:
$\exp[-\rm N_H(z)\, \sigma_{E[1+z]}]$;
(4) A redshifted Gaussian line, {\tt zgauss} to account for Fe-K$_{\alpha}$ emission, with the parameters describing the redshift, the energy of the line location, line width and the normalization defined as a total number of photons in the line in units of  $[\rm photons\,cm^{-2}\,s^{-1}]$;
(5) An {\tt apec} model describing emission from a collisionally-ionized plasma at a given temperature (in keV units), calculated from the AtomDB\footnote{http://atomdb.org/} atomic database \citep{Foster2012}, at the source redshift and assumed metallicity in terms of the solar abundances. 

We performed simulations to assess the significance of including the emission line and a thermal component using the {\tt{plot\_pvalue}} function in \sherpa\, \citep{Protassov2002}. Additionally, we ran the Metropolis-Hastings Bayesian Markov Chain Monte Carlo (MCMC) sampler with the {\tt{get\_draws}} function, assuming default flat priors and $5 \times 10^4$ iterations, to evaluate the posterior probability distributions for the complex models. We utilized the {\tt{corner}} Python package \citep{corner2016} to visualize these distributions (see Appendix~\ref{sec:Xray-Appendix}). 

We list the best-fit model parameters for the X-ray spectrum of \pks\ in Table~\ref{tab:model}, and present the results in Section~\ref{sec:X-ray}.

\section{Results}
\label{sec:X-ray}

\subsection{X-ray sources}

The centroid of \pks\ X-ray source is located at $\rm RA, DEC\, (J2000) = 0^h 25^m 49\fs24, -26^o 02' 12\farcs6$ with 0\farcs05 uncertainty.
\pks\ X-ray emission is not consistent with the emission of a point source. Figure~\ref{fig:profile} shows a surface brightness profile of \pks\ with an excess X-ray emission over the PSF profile at the radii $\sim 0.\arcsec4-5\arcsec$. There is no indication of an extended, X-ray diffuse emission surrounding the source on larger scales. We estimate the limit on the X-ray difusse emission below in Section~\ref{sec:xray-spectrum}.

We also detect a second X-ray source at $\rm (RA, DEC,J2000)= 0^h 25^m 50\fs56, -26^o 02'14\farcs7$, about 18$\arcsec$ distance to the East of \pks\ (see Figure~\ref{fig:field}). The source is present in the Gaia DR2 \citep{Gaia2018} optical catalog ID = 2323489174407213824 
and the average G magnitude of 
20.6525$\pm$0.0211. 
We discuss the X-ray properties of this source in the Appendix \ref{sec:Second-X-Appendix}.

\subsection{Distribution of the hot gas. } 
\label{sec:X-ray-image}

\begin{figure}
 \vskip 0.2in
   \centering
   \includegraphics[width=\columnwidth]{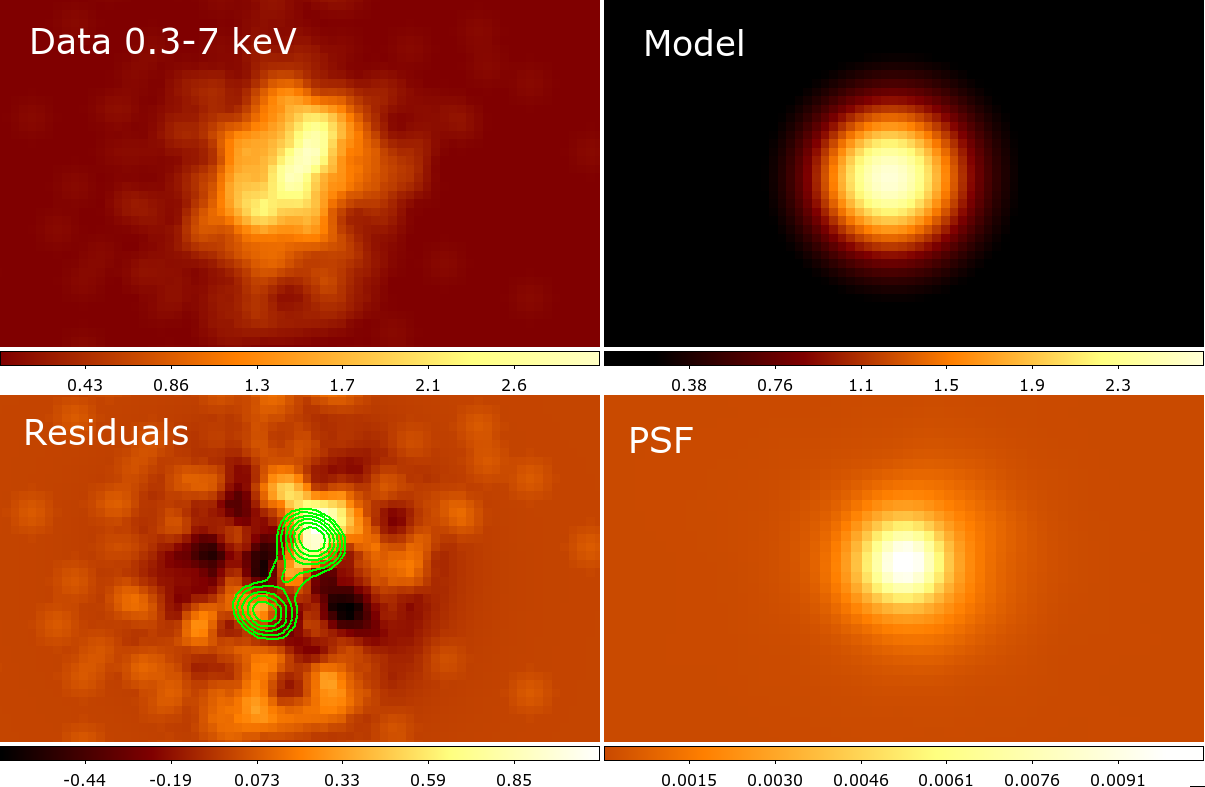}
\caption{2D model fit to the broad X-ray image of \pks. Four panels show the X-ray image (upper left), best-fit image of a 2D Gaussian circular model (upper right), the residuals (lower left) and the \chandra\ PSF image normalized to 1 (lower right). The color bars show intensity levels. The data and residual images were smoothed with the Gaussian with $\sigma=3$ pixels and the pixel size equal to 0.\arcsec123. The image of the residuals is overlaid with the 87\,GHz radio continuum contours with levels (1, 2.3, 5.6, 11, 19, 33) mJy/beam (beam = 0.\arcsec163$\times$0.\arcsec09474).
}
  \label{fig:image-fit}
    \end{figure}

We focus our analysis on the morphology of the hot gas in the central regions, within $< 10$\,kpc\ of the source.
In order to characterize the shape and size of the X-ray emission we fit the broad band X-ray image of \pks\ in \sherpa\ assuming a 2D Gaussian model convolved with the PSF. This model accounts for the AGN point source emission expected to follow the PSF shape and an additional smooth structure modeled as a Gaussian. 
Figure~\ref{fig:image-fit} shows the data, model, residual and the PSF images. The residual image shows
an elongated emission in the direction consistent with the radio source. Accounting for the ellipticity and rotation using the {\tt sigmagauss2d} model we obtained the deconvolved 1$\sigma$ width equal to $\sigma_{\rm maj} = 0\farcs36\pm0\farcs02$ and $\sigma_{\rm min}=0\farcs24\pm0\farcs02$. We note that this scale agrees with the distance between the peaks of the radio lobes of 0\farcs62 and may relate to the hot 'cocoon' emission surrounding the radio source. In this model there is still an excess of the X-ray emission towards the northern lobe indicating additional
enhancement possibly related to the jet-clouds interactions which we discuss in Section~\ref{sec:interaction}.

 \begin{figure*}
 \vskip 0.2in
   \centering
    \includegraphics[height=5.5cm]{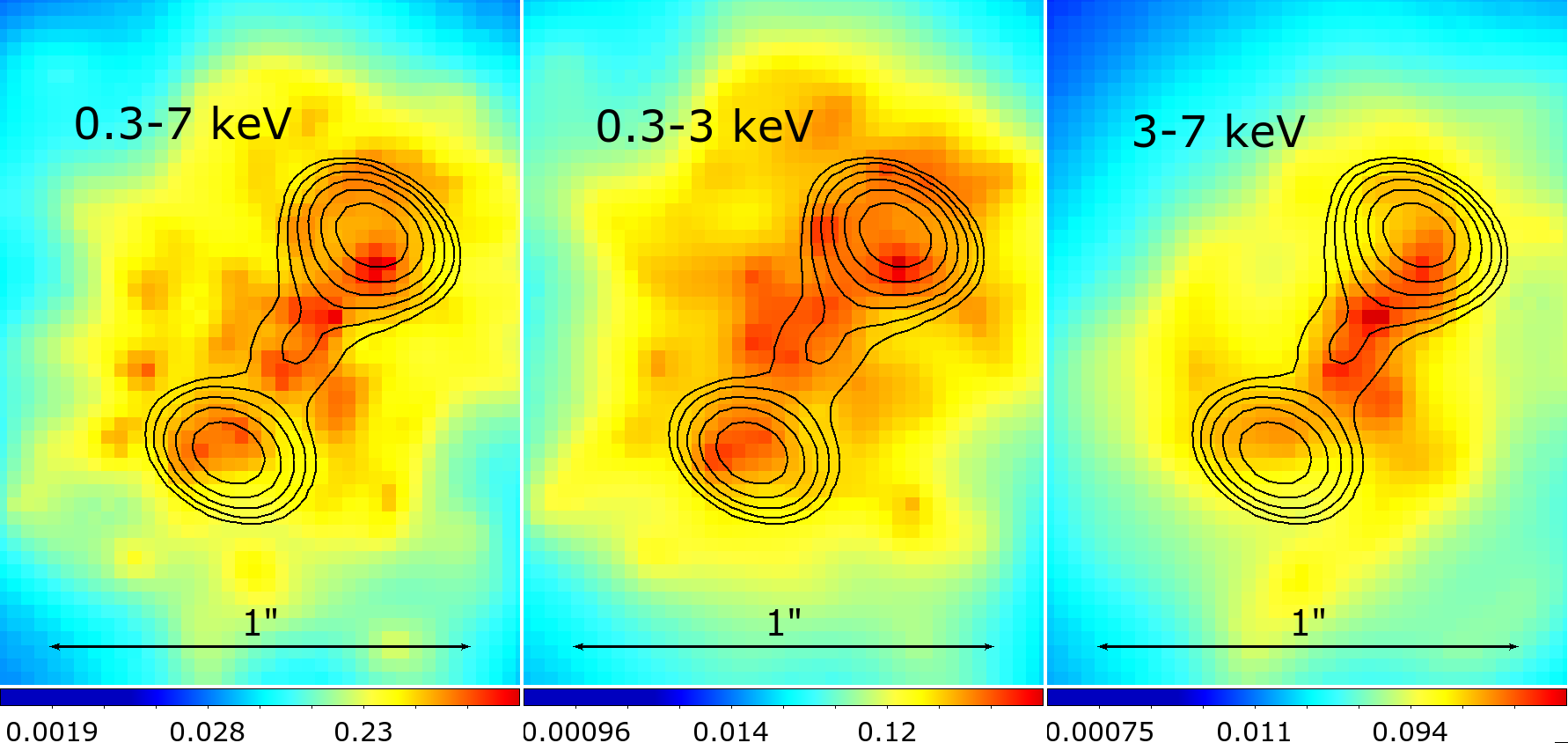}
\caption{\chandra\, image of \pks\, in three energy bands overlaid with the radio continuum contours. The X-ray images were binned to 1/16th ACIS pixel size and adaptively smoothed with a Gaussian kernel (see details in the text). The 87\,GHz radio continuum contours levels (are 1, 2.3, 5.6, 11, 19, 33) mJy/beam (beam = 0.\arcsec163$\times$0.\arcsec09474).}
  \label{fig:broad-soft-hard}
    \end{figure*}


 \begin{figure}
 \vskip 0.2in
   \centering
   \includegraphics[height=5.0cm]{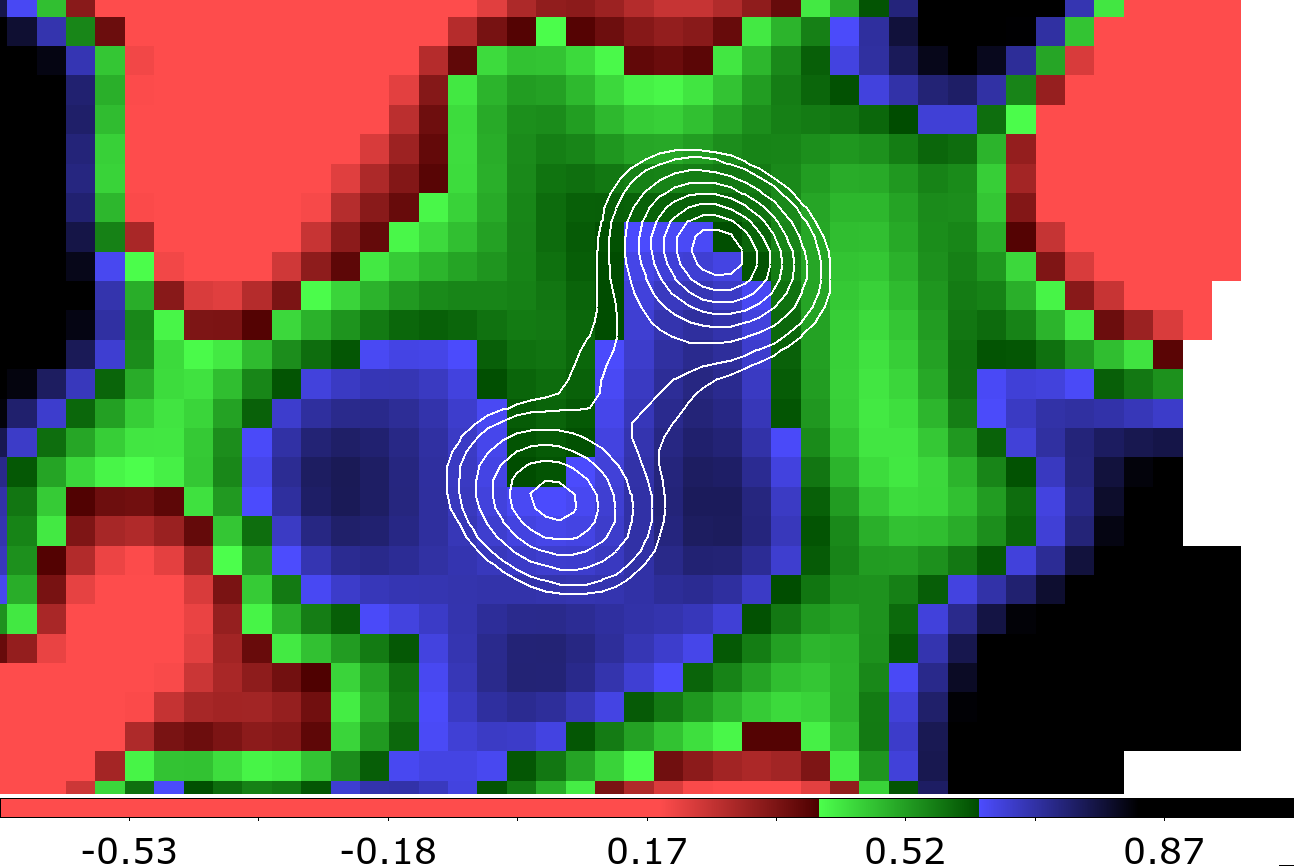}
\caption{X-ray hardness ratio color map overlaid with the contours of the radio source. 
The hardness ratio is defined as HR = H-S/H+S, where H = 2-7\,keV and S = 0.3-2\,keV.
The color scale is shown in the bottom of the image with the blue marking harder and the green/red softer X-rays. }
  \label{fig:hardness}
    \end{figure}

We generated the broad (0.3-7 keV), soft (0.3-3 keV) and hard (3-7 keV) band high resolution images by binning the merged event file into the 1/16th ACIS pixel size images which were adaptively smoothed with {\tt{dmimgadapt}} using Gaussian kernels ranging from 0.5 to 10 pixels (30 iterations and assuming 5 counts under the kernel).  Figure~\ref{fig:broad-soft-hard} shows the resulting smoothed images overlaid with the radio source contours. The X-ray emission is elongated, follows the direction of the radio source shown in contours, and has a spatial scale similar to the radio source. 
The soft band image shows a clumpy structure and enhanced emission at the location of the radio lobes. 
It is also slightly offset from the position of the radio core. The hard band X-ray radiation is more concentrated in the core region.

We further investigate the differences between soft and hard X-ray morphology using the hardness ratio (X-ray color) map defined as H-S/H+S, where H=2-7\,keV and S=0.3-2\,keV images. The resulting hardness ratio images are shown in Figure~\ref{fig:hardness} together with the radio source contours.
The harder X-ray emission surrounds the southern lobe and the western side of the radio source, although the overall radiation enclosing the radio source
is relatively hard (green and blue colors in the maps). 
Our spectral modelling of the X-ray emission in the smaller sub-regions also implies a difference between the southern and northern sub-regions which we describe in Section~\ref{sec:sectors}.

\begin{table*}
	\caption{Best-fit X-ray Model Parameters for the central r$\leq$1.5\arcsec\ region. }
    \vskip -0.3in
	\label{tab:model}
	\begin{center}
	\begin{tabular}{ c c c c c c c c c }
		\hline \hline
 \\
        Model &  N$_H(z)$  & $\Gamma$ & Line(E) or Line($z$) & EW (keV) & T(keV) & f$^{\,a}_{0.5-7.0 \rm keV}$  & Stat/d.o.f \\
\\
  \hline
  \\
 phabs*zphabs*pow  &  $<4.0$ & 1.23$^{+0.15}_{-0.12}$ & &  &   & 7.5$^{\pm0.6}$ & 475.8/443\\
 phabs*zphabs*(pow+zgauss) &  $<4.7$ & 1.32$^{+0.19}_{-0.13}$ & 6.32$^{\pm0.07}$ & 0.34$^{+0.16}_{-0.03}$  &  & 7.6$^{+0.7}_{-0.9}$ & 458.6/441 \\
 phabs*zphabs*(pow+zgauss)${^*}$ & $<4.9$& 1.31$^{+0.17}_{-0.13}$ & 0.339$^{\pm0.014}$  & 0.37$^{+0.01}_{-0.05}$ & & 7.5$^{\pm0.7}$ & 458.6/441 \\
 phabs*(zphabs*(pow+zgauss)+apec) &  6.1$^{+6.4}_{-5.2}$ & 1.45$^{+0.27}_{-0.25}$ & 6.32$^{\pm0.05}$  & 0.335$^{+0.11}_{-0.06}$ & 0.91$^{+0.25}_{-0.34}$ & 8.86$^{\pm0.03}$  & 449.2/439 \\
 phabs*(zphabs*(pow+zgauss)+apec)$^*$ &  6.1$^{+6.4}_{-5.2}$ & 1.45$^{\pm0.25}$ & 0.339$^{\pm0.010}$  & 0.331$^{+0.111}_{-0.096}$ & 0.91$^{+0.23}_{-0.33}$ & 9.38$^{\pm1.1}$  & 449.2/439 \\

\\

\hline
\hline
\\
		\end{tabular}
  
		Notes: Galactic absorption ({\tt phabs} model) of N$_H= 1.82\times 10^{20}$cm$^{-2}$ included in all models; N$_H(z)$ - intrinsic absorption ({\tt zphabs} model) is given in units of 10$^{21}$cm$^{-2}$; Redshift $z = 0.322$ is frozen for the intrinsic absorption and the thermal model ({\tt apec}). ${^*}$ - line energy is frozen at 6.4 keV ({\tt zgauss} model); Equivalent width (EW) and Temperature are given in units of keV. Abundances parameter is set to 1 in the {\tt apec} model. Uncertainties are 90\%; $^{a}$ Observed corrected for absorption flux in units of 10$^{-14}$$\rm erg\,cm^{-2}s^{-1}$;  
        The last column lists Cstat statistics and degrees-of-freedom for the best fit model. 
  \end{center}
\end{table*}

\subsection{Spectral Analysis and Results}
\label{sec:xray-spectrum}


\begin{figure}
   \centering
      \includegraphics[width=\columnwidth]{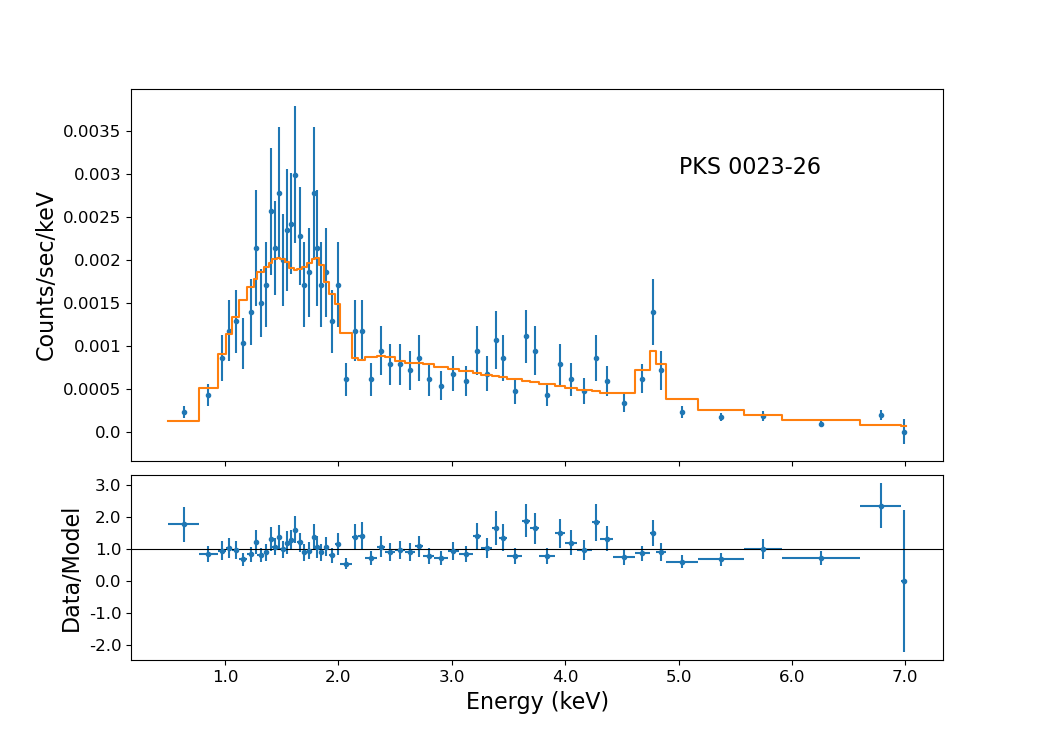}
\caption{\pks\ X-ray spectra fit with an absorbed power law model and addition of Fe-K${\alpha}$ emission line. The counts were grouped into 10 counts per bin for plotting. Top panel shows the data (blue points) and the model (orange solid line). The bottom panel displays the data to model ratio.}
\label{fig:x-ray-spectra}
\end{figure}
\noindent

We list the best fit parameters for the models applied to the X-ray spectrum of \pks\ in Table~\ref{tab:model}\footnote{Note that the X-ray spectrum was extracted from a circular region with the $r=1.\arcsec5$ radius.}. 
The best fit photon index of the power law models are relatively `hard', between 1.23$^{+0.15}_{-0.12}$ in the simplest model and 1.45$\pm0.25$ in the most complex one. We detect the 
Fe-K${\alpha}$ emission line (p-value $< 0.0002$), located at the rest frame energy of $\rm E_{rest}=6.34\pm0.07$\,keV and equivalent width of 0.34$^{+0.16}_{-0.03}$\,keV consistent with reflection from the cold medium \citep[e.g.,][]{Fukuzawa2011}. Figure~\ref{fig:x-ray-spectra} shows the absorbed power law model and Fe-K$\alpha$ emission line fit to the X-ray spectrum of \pks.

The most complex models (5 and 6 in Table~\ref{tab:model}) include
thermal emission 
from collisionally ionized medium ({\tt{apec}}) potentially associated with the jet-heated ISM \citep[see e.g.,][]{Fabbiano2019}. These are the most likely models for the X-ray spectrum of \pks. The MCMC sampling results show a relatively broad parameter distributions and a potential presence of a low temperature component (see Figure~\ref{fig:corner} in the Appendix~\ref{sec:Xray-Appendix}). However, we are  unable to evaluate the two-temperature model with the current data.
The {\tt{apec}} model with a plasma temperature of kT=0.91$^{+0.19}_{-0.37}$\,keV (1$\sigma$) is present in the X-ray spectrum (p-value$< 0.0002$) with 
the observed (0.5-2\,keV) flux equal to $1.1\pm0.3 \times10^{-14}$erg\,cm$^{-2}$\,s$^{-1}$, corresponding to the hot medium luminosity of $\rm L_{(0.5-2\,keV)} = 3.82 \pm 0.34 \times 10^{42}$erg\,s$^{-1}$.

The corrected for absorption flux of the power law model, in the 0.5-7\,keV energy band, is equal to 7.5$\pm 0.6 \times 10^{-14}\rm erg\,cm^{-2}s^{-1}$
which corresponds to the luminosity of 2.6$\times 10^{43}\rm erg\,s^{-1}$. 
We measure the luminosity of $3.1\times10^{43}\rm erg\,s^{-1}$ in the 2-10\,keV band, similar to the X-ray luminosity derived from XMM observations \citep{Mingo14}. We note that the XMM 30$\arcsec$ source extraction region contained the second X-ray source located 18$\arcsec$ away resolved in \chandra\, observations. However, this second source should not contribute to the XMM (2-10\,keV) luminosity as it has a softer spectrum and it is fainter than \pks\, in this energy band ($\Gamma=2.33^{+0.19}_{-0.28}$, 
$f^{\rm \, cont}_{(2 -10 \rm keV)} = 1.04^{+0.24}_{-0.17}\times 10^{-14}\, \rm erg~cm^{-2}~s^{-1}$, see Appendix~\ref{sec:Second-X-Appendix}).

We do not detect thermal emission on a large scale outside the central \pks\ region. However,
we can estimate the limit on this emission by modeling the X-ray spectrum from an r$\leq$12\arcsec\ circle, excluding the central r=1.\arcsec5.
For an absorbed power law with $\Gamma=1.45$ plus {\tt {apec}} model with kT=0.5\,keV we obtained 90\% limit on the flux of the apec model of $f^{\rm \, apec}_{(0.5 - 2 \rm keV)} < 9.5 \times 10^{-15}\, \rm erg~cm^{-2}~s^{-1}$, corresponding to the thermal luminosity of $ < 3.2\times10^{42}\rm erg\,s^{-1}$. For the higher temperature, kT=1\,keV, the limit is $f^{\rm \, apec}_{(0.5 - 2 \rm keV)} < 4.9 \times 10^{-15}\, \rm erg~cm^{-2}~s^{-1}$, corresponding to the luminosity of $ < 1.7\times10^{42}\rm erg\,s^{-1}$.
We note that this limit on the large-scale luminosity implies that the central X-ray emission cannot be associated with a cool core of a rich galaxy cluster (see discussion in Section~\ref{sec:GasOrigin}).

To summarize, our spectral analysis indicates the presence of thermal emission contributing to the central X-ray spectrum of \pks, and we are able to place a limit on the thermal emission at larger scale.  The primary AGN emission is absorbed and also reflected off the cold  medium as revealed by the Fe-K$\alpha$ emission line. The elongated shape of the X-rays closely aligned with the radio source indicates a jet-ISM connection and possibly suggesting jet heating. We will take this into account in the discussion in Section~\ref{sec:interaction}.

\subsection{Spectral Analysis of Sectors}
\label{sec:sectors}

\begin{figure}
   \centering
      \includegraphics[angle=0,height=5.0cm]{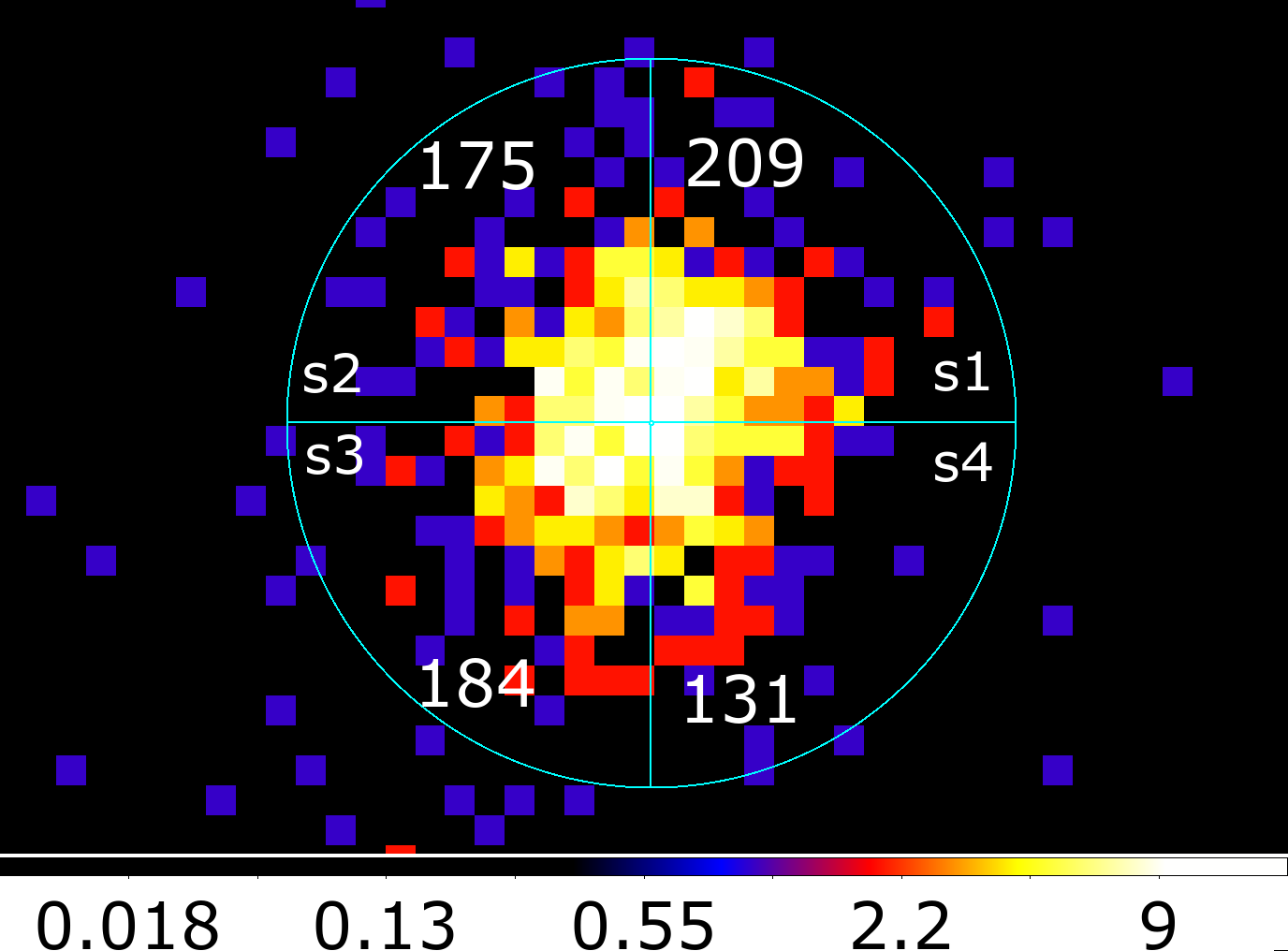}
\caption{X-ray image showing the pie regions selected for spectral analysis. The number of counts in 0.3-7\,keV energy range are marked in the image; the sectors are labeled as s1,s2,s3,s4. The X-ray image is binned into 1/4th ACIS pixel size (0.\arcsec123) and the outer radius of the circle is equal to 1.5$\arcsec$.}
\label{fig:sectors}
\end{figure}

We split the X-ray image along the circle with the radius of 1.5$\arcsec$ into four 90 degrees sectors, and obtained spectra for each sector (pie region in CIAO). Figure~\ref{fig:sectors} shows the regions with the corresponding number of X-ray counts in 0.3-7\,keV. The higher number of counts are detected in the regions 1 and 3, which also correspond to the north and south direction of the radio lobes. 

We applied an absorbed power law model plus a Gaussian emission line to the spectrum of each sector.
We also applied an absorbed {\tt apec} plasma model to each sector, however, we are unable to obtain valid solutions with the low counts spectra in these sectors.
We list the best fit model parameters in Table~\ref{tab:sectors}. Two sectors to the north, 1 and 2, show intrinsic absorption with the equivalent column density of about 5-6$\times 10^{21}$cm$^{-2}$, but with large uncertainties (upper bound in each case). There is no detectable absorption in the other two sectors and we only give upper limits. The Fe-K$\alpha$ emission line is strong in sectors 1,2,4 with a relatively high equivalent width and located at consistent energies, but the line is quite weak in sector 3. 

\begin{table}
	\caption{Best-fit X-ray Model Parameters for the Sectors.}
	\scriptsize
    \vskip -0.2in
	\label{tab:sectors}
	\begin{center}
	\begin{tabular}{ l c c c c c c c c }
		\hline \hline
 \\
        Parameter & s1  & s2 & s3 & s4 & \\
        \\
\hline
  \\
  $N_H(z=0.322)$ & $5.2 (< 7.0) $  & 5.9 ($<8.8$) & $<3.2$ & $<2.5$ \\
  $\Gamma$  & 1.77$^{+0.47}_{-0.43}$ &  2.24$^{+0.63}_{-0.52}$ &  1.15$^{+0.26}_{-0.23}$ & 1.20$\pm0.27$ \\
  Norm & 3.8$^{+2.9}_{-2.5}$ & 4.7$^{+4.9}_{-2.1}$ & 2.1$^{+0.7}_{-0.5}$ & 1.67$^{+0.56}_{-0.41}$ \\
  E(Fe-K$\alpha$) & 6.28$\pm0.12$ & 6.22$^{+0.19}_{-0.21}$ & 6.4$\pm0.5$ &  6.33$\pm0.08$ \\
  EW & 0.450$^{+0.374}_{-0.267}$ & 0.742$^{+0.662}_{-0.336}$ &  0.204$^{+0.174}_{-0.155}$ &  0.735$^{+0.119}_{-0.400}$  \\
  \\
\hline
\\
		\end{tabular}
        
		Notes: Galactic absorption ({\tt phabs} model) of N$_H= 1.82\times 10^{20}$cm$^{-2}$ included in all models; N$_H(z)$ - intrinsic absorption ({\tt zphabs} model) is given in units of 10$^{21}$cm$^{-2}$; Redshift $z = 0.322$ is frozen for the intrinsic absorption. The normalization of a power law model is defined at 1 keV and given in the units of $10^{-6}$ photons~cm$^{-2}$~s$^{-1}$~keV$^{-1}$. 
        Equivalent width (EW)  in keV units. 
        Uncertainties are 90\% except for EW with 1$\sigma$ errors from simulations. 
      \end{center}
\end{table}

\section{Discussion}
\label{sec:discussion}

We presented the results from a deep \chandra\ observation of a powerful radio source, \pks. We found an absorbed X-ray nucleus with the Fe K-$\alpha$ emission line that signals the presence of a reflection medium within the radius $<7$\,kpc of the X-ray source region.
Our high angular resolution X-ray image traces the distribution of hot gas which is closely aligned with and extends beyond the radio source, especially  in the direction perpendicular to the radio source axis (soft X-rays). 
We did not detect any diffuse X-ray emission outside the radius of $\sim$20\,kpc and concluded that despite several galaxies present at similar redshift there is no rich X-ray cluster associated with \pks. We placed a 90\% limit of $L_{(0.5-2\rm keV)} < 3.2\times 10^{42}$\,erg\,s$^{-1}$ on the luminosity from a hot plasma with a temperature of kT=0.5\,keV, which is consistent with X-rays from a group of galaxies. 
In the following, we discuss our results in the context of an evolving radio jet and its impact on the ISM.

\subsection{Large Scale Environment}
\label{sec:GasOrigin}

Based on galaxy number counts analysis, \pks\ appears to be located in a relatively rich environment, with a number of close companions seen in projection \citep{Ramos13}.
Three of these companions -- all within a projected radial distance of 120\,kpc  of \pks\ -- have spectroscopic redshifts that are similar to \pks\ (\cite{Tadhunter2011}: 
radial velocity differences of
+690$\pm$34, +200$\pm$75 and +280$\pm$60 km s$^{-1}$ relative to the rest frame of \pks). 
However, our \chandra\ X-ray image shows that the emission is concentrated on \target\ and a second object about 18\arcsec\, away (i.e. $\sim 80$ kpc) but does not show the X-ray diffuse component typically associated with clusters of galaxies. While the X-ray luminosity of the gas component in the central 1.5 arcsec ($\sim$7~kpc) is relatively large ($\sim$3$\times 10^{43} \rm erg\,s^{-1}$) the limit on gas luminosity in the surrounding 12\arcsec\ (56.6~kpc) is much more modest (a few $\times$10$^{42} \rm erg\,s^{-1}$).

Comparison with the X-ray luminosity profiles of other nearby systems confirms that if \target\ were located at the center of even a poor galaxy cluster \citep[e.g., AWM~4,][]{OSullivanetal10} or a rich, cool-core galaxy group \citep[e.g., NGC~5044,][]{OSullivanetal17} the extended intra-cluster or intra-group medium (IGrM) would be visible in our 1.5-12\arcsec-radius \chandra\ spectrum. However, the IGrM of a less massive group could go undetected; systems with typical temperature $\sim$0.5~keV can have luminosities of only a few $\times$10$^{41} \rm erg\,s^{-1}$ within 50~kpc \citep{Helsdonetal05}.

The distribution of the cold molecular gas observed by ALMA  (\citealt{Morganti2021,Oosterloo2025}, Paper 1)
suggests the presence of on-going interaction between the host galaxy of \target\ and surrounding galaxies,  with tails of gas pointing in the direction of nearby (in projection) galaxies.
The origin of the large amount of cold molecular medium are likely to be due to accretion from these companions. This is in contrast to cool-core clusters (e.g., \citealt{McNamara2016,Russell2019}). Cooling from the IGrM can be a source of molecular gas in the dominant ellipticals of galaxy groups with X-ray luminosities comparable to the upper limit on extended emission around \target\ \citep{OSullivanetal18,Olivaresetal22}. However, examples of tidal accretion of cold gas from neighbours become more common in lower-mass galaxy groups \citep[e.g., in NGC~5903, NGC~315, and NGC~1587;][]{Appletonetal90,Morganti2009,Olivaresetal22}, unsurprisingly since the group environment brings galaxies into close proximity at low relative velocities, and galaxies are much less likely to be ram-pressure stripped than they would be in a rich cluster.

In central cluster galaxies (BCG) the observed molecular gas structures seen around the radio bubbles (which contain radio jets) are explained as either a thin cover of clumpy molecular gas, expanding along with the raising radio bubbles, or molecular gas that is in situ condensed in the updrafts (e.g. \citealt{McNamara2016,Russell2019}). In \target\ the former hypothesis appears more likely: what we are observing is the interaction between the expanding/growing radio lobes and the molecular gas accreted from companion galaxies. 

Past studies of quasars indicated no differences between the environment of radio-quiet and radio-loud quasars, with only a few radio quasars found in X-ray bright clusters. Recent studies of the environment  of low redshift radio galaxies and radio quasars using a low frequency radio observation in
the LOFAR Two-meter Sky Survey (LoTSS DR2) \citep{Pan2025} confirmed that luminous quasars are very rare in clusters, whereas a large fraction of less luminous radio galaxies are located in cluster centers. We find that \pks\ is in a group (not strong cluster) which agrees with this statistical finding and also supports our findings about the origin of the molecular gas in \pks.

\subsection{Molecular Clouds Environment}
\label{sec:comparison}

In the companion paper (\citealt{Oosterloo2025}, Paper 1) we use ALMA observations of several CO transitions to study the morphology, kinematics and physical conditions of the molecular gas.
On large scales, the CO distribution is uniform and appears quiescent with a few streams of molecular gas extending towards nearby galaxies.  
The conditions in the vicinity of the radio source are quite different.
We found extreme physical conditions implied by a non-detection of the \coOne\, gas in the radio core region, where the peak of the \coThree\ gas distribution is located. 
As shown by Fig.~\ref{fig:broad-soft-hard} this is also the region where the X-ray emission is most intense. 
\citealt{Maloney1996} considered physical processes in X-ray irradiated molecular gas and showed that X-ray ionization can lead to destruction of CO molecules.
A lack of \coOne\ line emission coincident with intense X-ray radiation fields has been reported previously \cite[e.g.][in X-ray clusters]{O'Dea1994}. In ESO 428-G014 galaxy, there is a lack of \coTwo\
emission in the nuclear region, where instead molecular gas emitting H2 is abundant \citep{Feruglio2020}, leading to the suggestion by these authors that the X-ray photons may interfere with the production of \coOne.  
They suggested that CO may be excited to higher excitation levels up the CO ladder in this region, due to irradiation by hard X-ray field and/or shocks. A similar effect was noticed in NGC 2110 \citep{Fabbiano2019c}, where X-ray emission was detected with \chandra\ to fill the \coOne\
nuclear 'lacuna' \citep{Rosario2019}.
The lack of \coOne\ 
emission in the nuclear region of \pks, contrasted with the X-ray emission peak in the same region, is consistent with these previous suggestions and theoretical predictions that the X-ray photons may excite CO to higher excitation levels in regions of intense X-ray photon fields. The close alignment between the molecular gas distribution, elevated CO excitation levels, and the radio axis further indicates that these intense X-ray fields likely result from interactions along the jet 

\begin{figure}
   \centering
      \includegraphics[angle=0,height=5.5cm]{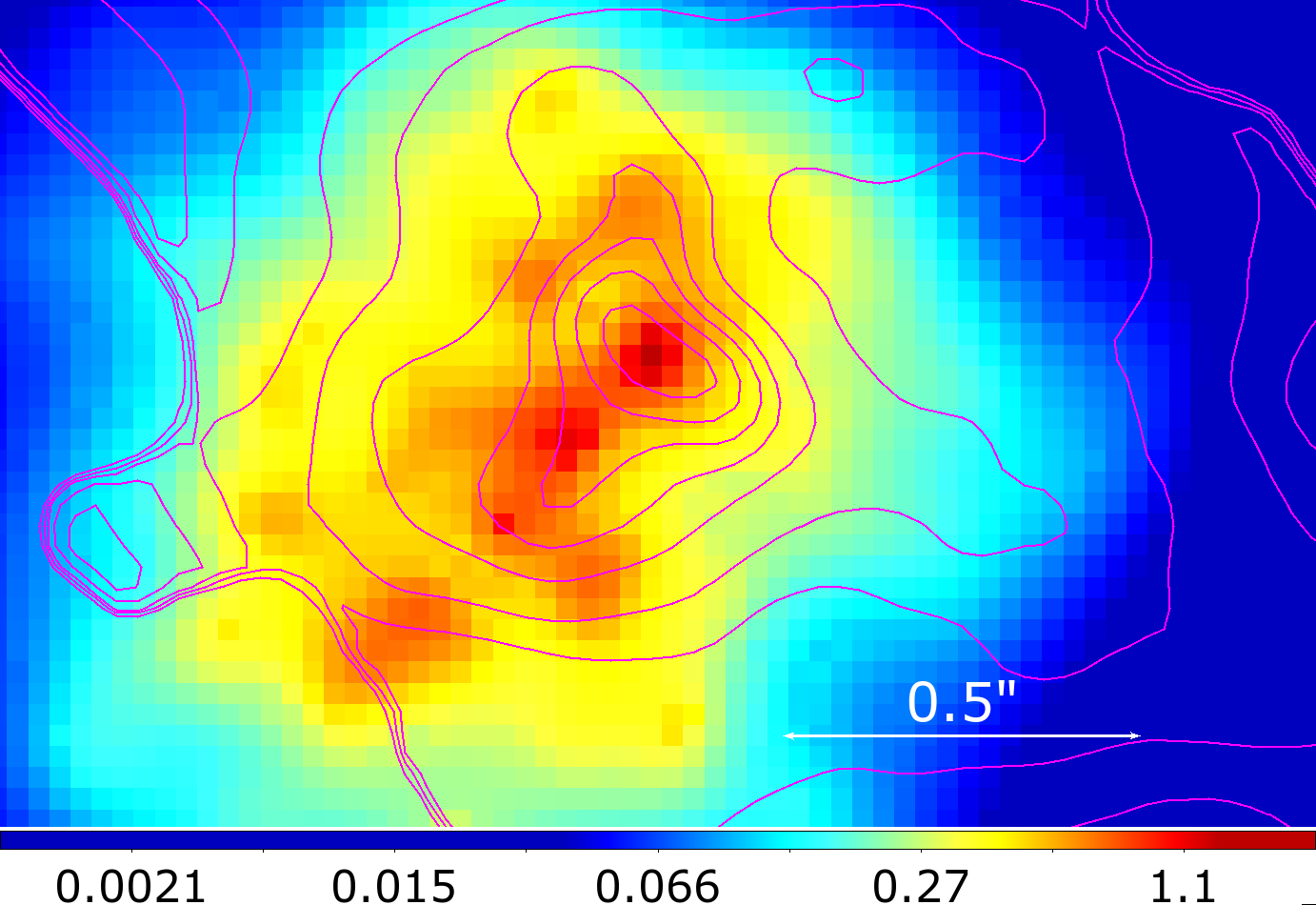}
\caption{X-ray image (0.3-7 keV) overlayed with the contours of the CO(3-2)/CO(2-1) line ratio from the ALMA observations (\citealt{Oosterloo2025},Paper 1).
The X-ray image was adaptively smoothed as in Figure~\ref{fig:broad-soft-hard}. The magenta shows the contours from the line ratio map with the levels starting at the peak 1.05, 1.0, 0.95, 0.9, 0.82, 0.7, 0.6, 0.4, 0.2}.
\label{fig:CO-ratio}
\end{figure}

In ALMA data we also observed a large velocity dispersion of cold gas at the northern radio lobe, suggesting interactions between the expanding radio lobe and the molecular ISM, inducing turbulence (\citealt{Oosterloo2025}, Paper 1). Such interactions were also supported by the detection of high \coThree\ to \coTwo\ line ratios, \rdt\ (Fig.~\ref{fig:CO-ratio}). 
At the edge of the radio lobes, about $\sim 2$\,kpc from the core, the molecular gas traced by \coTwo\ and \coThree\
appeared to wrap around the radio source where the molecular gas was pushed aside or blocked from reaching the central regions. Furthermore, the \coTwo\ and \coThree\ radiation appeared to be faint in the regions of the radio lobes. Interesting, the \coThree\ to \coTwo\ line ratio map (Fig.~\ref{fig:CO-ratio}) indicates that the X-ray emission at the site of the northern lobe coincides  with the region of the highest line ratio (dominated by the \coThree), 
supporting the idea of the jet driving strong shocks through dense clouds, heating and destroying them, and producing  thermal X-ray radiation.

The distribution and kinematics of the molecular gas at the end of the jet suggest that the interactions compress the ISM and increase the density resulting in efficient cooling. Such interactions may perhaps explain the morphology of the molecular gas around the northern radio lobe. This point might also be supported by the X-ray hardness ratio map (Fig.~\ref{fig:hardness}), showing softer X-rays in the northern regions, similar to what was observed at the location of the jet-ISM interaction in another radio galaxy NGC1167/B2\,0258+35 \citep{Fabbiano2022}.

\subsection{Jet-ISM interactions - Heating and the Cocoon}
\label{sec:interaction}

The distribution of CO line ratios outside the radio source, in particular in the region perpendicular to the jet direction, implying the presence of a cocoon of gas resulting from the jet-ISM interactions (Oosterloo et al. 2025).
Our \chandra\ images show the X-ray emission elongated in the same direction as the radio emission, and enveloping it, suggesting jet heating (see Sect.\ \ref{sec:xray-spectrum}) 
and supporting the presence of a cocoon of shocked gas.
Therefore, we might expect that the X-rays are produced by strong interactions between the
jet and the pre-existing warm/cold gas: as the gas structures are shredded and heated by the jet-induced shocks, perhaps they could produce X-ray emission that is enhanced compared to that of the 
underlying X-ray halo (e.g. in 3C171, \citealt{Hardcastle2010a}: 3C305, \citealt{Hardcastle2012}; PKS B2152-69, \citealt{Ly2005,Worrall2012}). If the pre-existing warm/cool medium has fairly low 
densities (perhaps not too unlikely on kpc scales) it may not be able to survive the shredding to cool 
to a molecular phase, hence explaining the  hollowing out of the molecular gas at the sites of the lobes.

We can place some limits on the mass of hot gas in the cocoon based on its X-ray luminosity, but these depend on the volume assumed. If we assume a prolate ellipsoid cocoon closely coupled to the radio jets, then the X-ray image suggests that its length is $\sim$1.13\arcsec\ (5.4~kpc, along the jet axis) and width $\sim$1\arcsec\ (4.7~kpc). Subtracting the volume of the radio lobes (considered as 1.6~kpc diameter spheres), the X-ray emission would arise from a volume $\sim$52.4~kpc$^3$, containing $\sim$3.8$\times$10$^8$~M$_\odot$ of hot gas. This is comparable within a factor of a few with the hot gas masses found in the central few kiloparsecs of group dominant galaxies \citep[e.g., 10$^8$~M$_\odot$ in the central 2.4~kpc of NGC~777,][]{OSullivanetal24}, so it is plausible that a significant fraction of the cocoon material may arise from a pre-existing hot medium. However, we note that this mass estimate is relatively conservative, and will increase if the volume is larger than we have assumed; in the unlikely case that the hot gas fills the whole 1.5\arcsec-radius spectral extraction region the mass would be $\sim$2$\times$10$^9$~M$_\odot$, much less consistent with the hot gas masses found in group cores. In either case, a contribution from shock-heated low-density cold gas may also be needed. We note that some of the hot gas mass in the aligned structures might originate in the destruction of the cold molecular clouds by the interaction of the radio jets with the cool, merger-origin ISM in the central parts of the galaxy.

\subsection{Nuclear Region}

\label{sec:reflection}

The results of the X-ray spectral analysis of \pks\, presented in Sec.~\ref{sec:xray-spectrum} were based on the source area covering the central $<7$\,kpc region of the host galaxy which include the radio source and molecular gas resolved by ALMA. 
We detect intrinsic absorption with an equivalent neutral hydrogen column density of $6.1^{+6.4}_{-5.2}\times 10^{21}$\,cm$^{-2}$, the spectrum with a relatively hard photon index of $\Gamma=1.45\pm0.25$ (uncertainties are 90\%) and Fe-K-$\alpha$ emission line.

The Fe-K-$\alpha$ emission line with the $EW=0.335^{+0.11}_{-0.06}$\,keV could originate anywhere within this region. The fluorescence Fe-K-$\alpha$ line is usually interpreted as a reflection of the cold medium in the inner few parsecs region around SMBH. However, recent \chandra\ studies of several nearby galaxies found hard X-rays and Fe-K-$\alpha$ emission on larger kiloparsec scales \cite[e.g.][]{Fabbiano2018,Fabbiano2019b,Jones2020,Jones2021}. Our high resolution image analysis show slightly harder emission towards the southern radio lobe. The spectral modeling of the sectors shows differences in the equivalent width of the emission line potentially indicating the preference for the reflection angles in the sectors perpendicular to the radio axis.

We measure the corrected for absorption X-ray luminosity of $\rm L_{2-10\, \rm keV} = 3.1 \times 10^{43}$~erg\,cm$^{-2}$s$^{-1}$ which is in agreement with the earlier \xmm\ measurements by Mingo et al 2014. However, we note that the spectral energy range in the \chandra\ observation limits our determination of the intrinsic emission of the AGN and additional hard X-ray data, at $>$10keV, are needed to improve current constraints on luminosity and absorption.

\subsection{Outflows}
\label{sec:outflow}

Multi-band observations show different types of outflow present in
\pks. The high jet power (few $\times 10^{46}$ \ergs) derived by \cite{Morganti2021} indicates that the radio source and the jet have enough energy on kiloparsec-scales to accelerate and disperse the gas. Strong AGN radiation could also power the outflows \citep{Begelman1983, Harrison2018}. 
An outflow of ionized gas with a mass outflow rate in the range $0.13<\dot{M}<1.47$ \msunyr was indicated by strong and broad (${\rm FWHM}>1000$ \kms) optical emission lines 
\citep{Holt2008,Shih2013,Santoro20}. 
The ionized outflow has a radial extent larger than the size of the radio source and exhibits high densities attributed to gas compression resulting from the jet-ISM interaction. However, the strong emission lines are attributed to AGN photoionization and not shocks associated with the interactions \citep[see][]{Santoro20}. 
The mass outflow rate of the cold molecular gas
within the $\sim$1\,kpc region observed with ALMA is $\dot{M} \lesssim  $ 20 \msunyr. 
This is much higher than the outflow rate of the ionized gas component (\citealt{Santoro20}) which agrees with the outflow rates observed in other sources. 

X-rays could arise from AGN photoionized gas or shock heated medium.
\chandra\ studies of low redshift radio galaxies where the soft X-rays spectra could be spatially resolved imply the mixture of photoionized and shock heated gas present in these systems (e.g. \citealt{Wang2011}, D.\L. Kr\'ol et al. 2025). \chandra\ observations of powerful radio galaxies with aligned radio and X-rays morphology, and the spatial-spectral properties of hot gas points to the shock heating process \cite[e.g. 3C171, 3C305][]{Hardcastle2010a,Hardcastle2012}. 
In \pks\
the X-ray emission is closely aligned with the radio axis and has a similar extent to the radio source in the direction of the radio axis which also argues for the shocks.

Measurement of the outflow rate of the hot gas in \pks\ is challenging due to the low signal of our data.  Although, we can estimate the upper limit of the outflow rate assuming that the entire hot gas with a mass of $3.8 \times10^8 M_{sun}$ (see Sec.\ref{sec:interaction}) participates in the outflow, such a scenario is unlikely.
Using, $\dot M \sim M_{out} \times v_{out}/r_{out}$, and the same outflow velocity as the velocity of the ionized wind, $v_{out}=890$\,km/s \citep{Santoro20}), 
and the extent of $\sim 1$\,kpc we get $\sim 350 \, M_{\odot} \rm yr^{-1}$. This is a very high mass outflow rate in comparison to the rates typically observed in X-ray outflows \citep[see e.g.,][]{Fiore2017}. On the other hand, the recent observation of a radio-quiet quasar, PDS\,456, shows such high mass outflow rates with relativistic velocities of the clumpy wind \citep{Xrism2025}. 

We can calculate the sound speed in hot uniformly distributed gas using the expression $c_s=\sqrt{\hat \gamma kT / \mu m_p}$, assuming non-relativistic gas ($\hat \gamma ={ 5\over3}$) and the mean molecular weight of the gas $\mu=0.6$. For the gas temperature of kT=0.9\,keV the measured sound speed is $\sim 500$\, km/s. 
Given the high velocity of the warm outflows ($\sim 1000$\,km/s), which exceeds the local sound speed, the presence of shocks is inevitable. If we assume that temperature kT=0.9\,keV is the post-shock temperature then the expected shock velocity for fully ionized gas would be $v_{shock} =\sqrt{16  kT / \mu m_p} \sim 870$\,km/s, so relatively close to the velocity of warm outflow. The Mach number for such shock would be within $\mathcal{M} \sim 1.75 - 2$, so the shock would be in the supersonic regime with the gas compressed by a factor of $\sim 2$.
Although, such shocks cannot be directly resolved in our X-ray data, our observations support the presence of a cocoon structure resulting from the interaction between the expanding radio source and the surrounding medium.

Similar and also stronger shocks driven by radio source expansion into the interstellar medium (ISM) have been resolved in other systems. For example, shocks with Mach numbers of $\sim 4$ have been detected within $\sim 10$\,kpc in the radio galaxy NGC 3108 \citep{Croston07} and the Seyfert galaxy Markarian 6 \citep{Mingo2011}. Additionally, a weaker shock with a Mach number $<2$ has been observed at the site of a radio lobe approximately $\sim 20$\,kpc from the galaxy center \citep{Siemiginowska2012}.

Strong outflows are not commonly associated with galaxies in the centers of galaxy clusters. 
Furthermore, quasars residing within X-ray clusters are extremely rare with only a few currently known cases \citep[e.g.,][]{Siemiginowska2005, Siemiginowska2010, Russell2010}. Therefore perhaps a non-detection of strong X-ray emission typical of a galaxy cluster (see Section~\ref{sec:GasOrigin}) is not surprising.

\section{Summary and conclusions} 
\label{sec:conclusions}

We discussed the results of a deep Chandra X-ray observation of a young radio source \pks\ associated with a powerful quasar. The source is surrounded by cold molecular gas studied with ALMA. Several galaxies with similar redshift indicated a possibility that the diffuse X-ray emission from a cluster of galaxies can be observed in X-rays. 
We did not detect any diffuse X-ray emission at the luminosity expected from a typical X-ray cluster. We placed a limit on the X-ray luminosity of a few $\times 10^{42}$\,erg\,s$^{-1}$ within $\sim60$\,kpc region located outside the central $r=7$\,kpc region. This limit does not rule out a possibility that \pks\ is located in a low temperature poor galaxy group.

We observed the X-ray morphology of hot gas in the central region to be elongated and extended beyond the radio source in the direction perpendicular to the jet axis. The enhanced X-ray emission at the location of the northern lobe points to the site of jet-clouds interactions resulting in shock heating of the gas. 
It is also the location of the peak of the \coThree/\coTwo\, line emission, suggesting that the interactions between the jet and cold medium result in the X-ray radiation which 
excites CO.
The estimated Mach number of $\mathcal{M} = 1.75-2$ for the shock in \pks\, is in agreement with observations of shocks in other radio galaxies pointing to a prevalent impact of jets on ISM.

The X-ray spectrum of the central region ($r=7$\,kpc) shows a mildly absorbed AGN nucleus and Fe-K$\alpha$ emission line at $E\sim6.4$\,keV signaling a reflection from the cold gas. The observed equivalent width of the line is in agreement with mildly absorbed AGN. However, the limited spectral energy coverage does not allow for detailed constraints of the intrinsic AGN spectrum and the amount of absorption. The X-ray spectra at higher energy, such as provided by NuStar are needed to better constrain these parameters in the future.

\begin{acknowledgements} 

We thank the referee for comments and additional references which improved the manuscript.
We thank Dominika Kr\'ol for a discussion of the otflows ionization structure, Beatrice Mingo for a discussion of the XMM data and Kenny Glotfelty for helping with merging Chandra observations.
Support for this work was provided by the National Aeronautics and Space Administration through Chandra Award Number GO1-22118X
issued by the Chandra X-ray Observatory Center, which is operated by the Smithsonian Astrophysical Observatory for and on behalf of the National Aeronautics Space Administration under contract NAS8-03060.
A.S., G.F., E.O. and D.B. acknowledge support from NASA Contract NAS8-03060 to the {\sl Chandra X-ray Center}.

\medskip
This research has made use of data obtained from the Chandra Data Archive, and software 
provided by the Chandra X-ray Center (CXC) in the application packages CIAO and Sherpa.
It employs a list of Chandra datasets, obtained by the Chandra X-ray Observatory, contained in the Chandra Data Collection (CDC)~\dataset[https://doi.org/10.25574/cdc.400]{https://doi.org/10.25574/cdc.400}

This paper makes use of the following ALMA data: ADS/JAO.ALMA\#2022.1.01498.S. ALMA is a partnership of ESO (representing its member states), NSF (USA) and NINS (Japan), together with NRC (Canada), MOST and ASIAA (Taiwan), and KASI (Republic of Korea), in cooperation with the Republic of Chile. The Joint ALMA Observatory is operated by ESO, AUI/NRAO and NAOJ. 
\end{acknowledgements}

\clearpage
\newpage

\begin{appendix}

\section{\pks\, X-ray Data Analysis - additional figures.}
\label{sec:Xray-Appendix}

   \begin{figure}[h]
   \centering
\includegraphics[width=0.5\columnwidth]{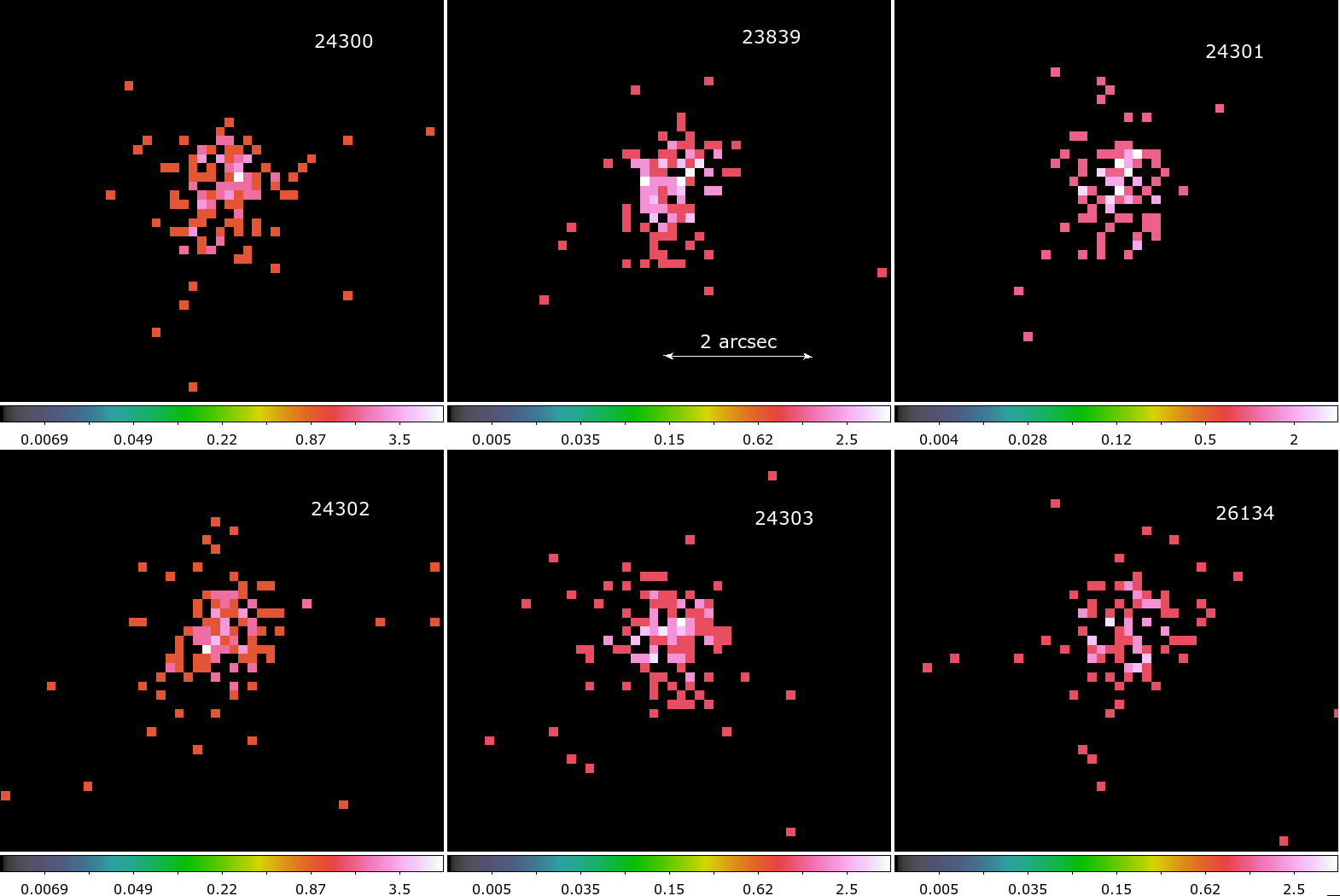}
   \caption{Chandra ACIS-S X-ray image of \pks\ in the 0.3-7\,keV energy range. Each panel is $5.\arcsec2 \times 6.\arcsec0$ in size and displays a single observation with the obsid number marked in the upper right corner.  The  2\arcsec\ scale bar is shown in the upper central panel. The pixel size is 0.123\arcsec. The color indicates number of counts per pixel and the color bar is shown at the bottom of each panel. 
   }
               \label{fig:x-images}
    \end{figure}


 \begin{figure}
 \vskip 0.2in
   \centering
\includegraphics[width=0.5\columnwidth]{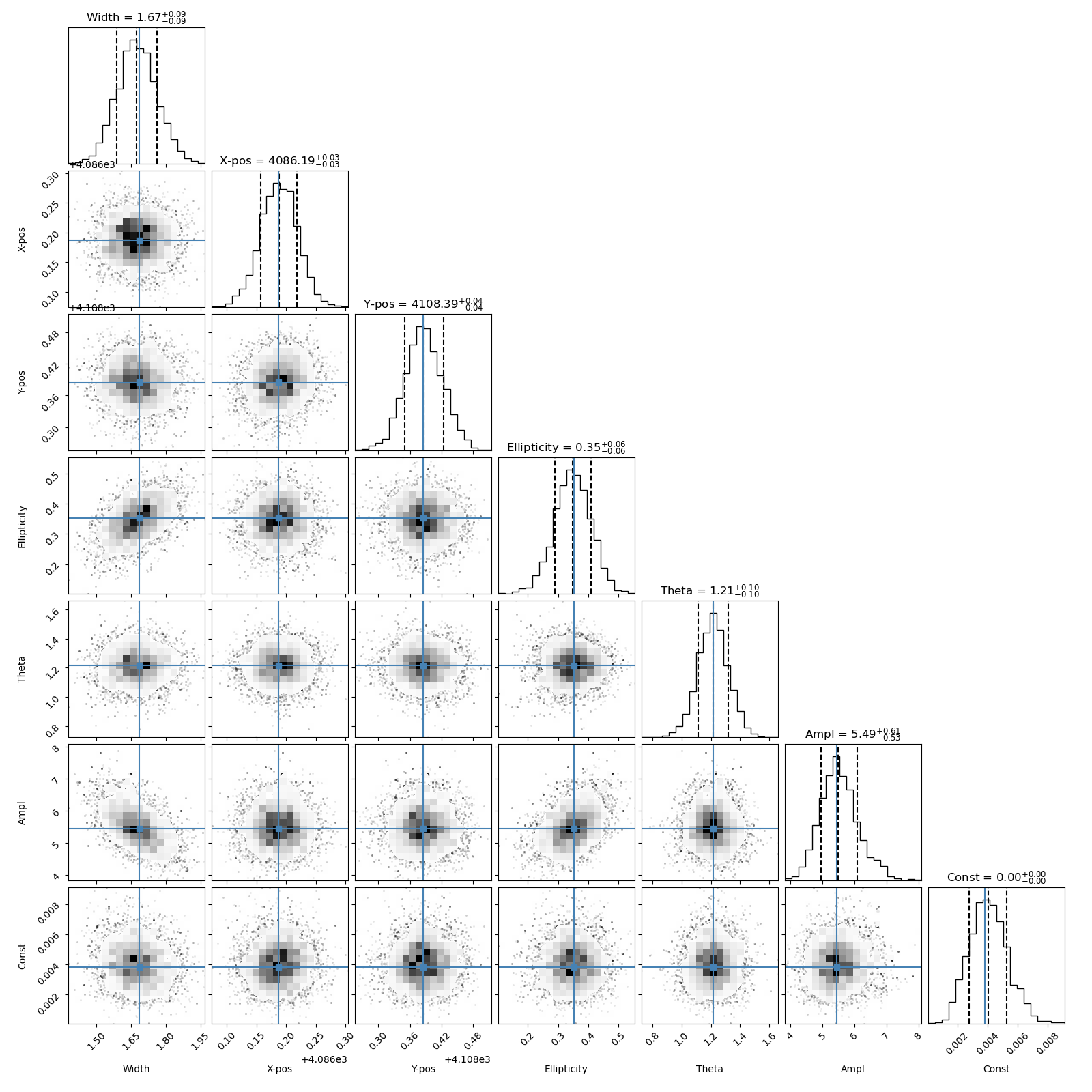}
\hskip 1in
\caption{The parameter distributions for the 2D Gaussian model fit to the broad X-ray image of \pks\, obtained using {\texttt{get\_draw()}}, the Bayesian MCMC sampling in Sherpa. 
Model expression as defined in the Sherpa fit: {\texttt{psf(gauss2d+const2d)}}. The best fit model parameters are listed above relevant histograms and are marked by blue vertical lines, the mean and 1$\sigma$ range for the distribution is marked by dashed vertical lines. The corner plot was made using {\tt corner} python package 
\citep{corner2016}.
}
  \label{fig:image-fit-corner}
    \end{figure}


   \begin{figure}
   \centering
\includegraphics[width=0.5\columnwidth]{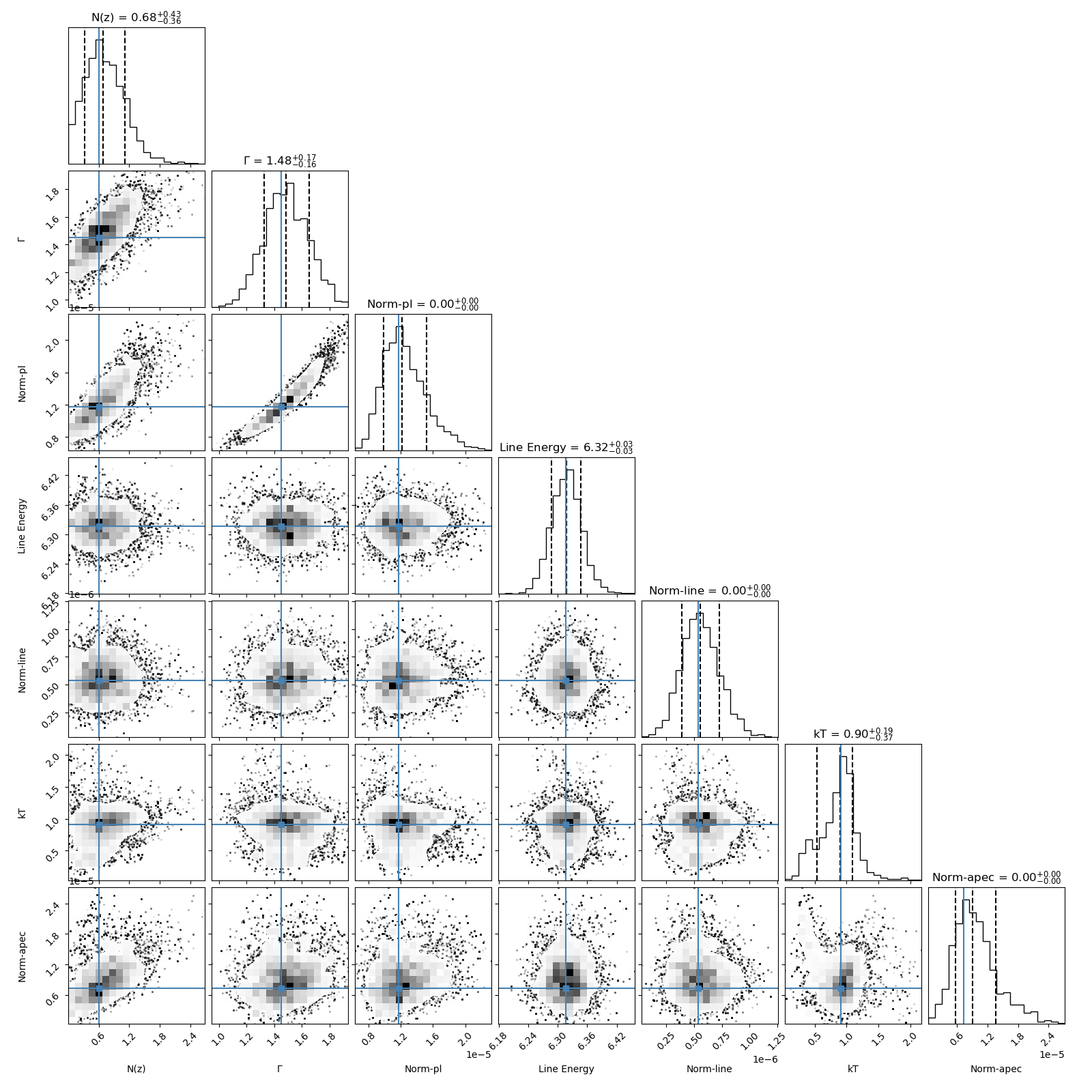}
\caption{Distributions of the model parameters for spectral model fit to \pks\, {\tt{phabs*zphabs*(powlaw1d+zgauss)+phabs*apec}} from the Metropolis-Hastings sampling with {\tt{get\_draws}} in Sherpa. The labels show parameters on each axis. The best-fit values from the fit are marked by solid blue square and lines, while the distribution quantiles are marked by dashed lines, the center dashed line shows the mean. The top of each column shows the mean of the parameter value and 1$\sigma$ quantiles.}
    \label{fig:corner}
    \end{figure}

\begin{table}
{
\caption{\label{tab:fit-obsids} Model Parameters for Single Epoch Observations}
\begin{center}
\hfill
\vskip -0.1 in
\begin{tabular}{cccrcrrl}\hline\hline
 Date & obsid$^a$  & Photon Index &  Norm$^b$ & f$^c_{0.5-7.0 \, \rm keV}$ & cstat/d.o.f. \\
\hline
2021-09-13 & 24302 & 1.3$^{+0.34}_{-0.28}$ & 10.0$^{+4.9}_{-2.5}$  & 7.6$^{+3.5}_{-0.3}$ & 305.8/444\\
2021-09-19 & 24301 & 0.94$^{+0.31}_{-0.32}$ & 8.7$_{-2.6}^{+3.4}$   &  9.5$^{+6.5}_{-3.8}$ & 296.4/444\\
2021-09-20 & 23839 & 0.97$^{+0.27}_{-0.27}$ & 8.7$_{-2.3}^{+2.8}$   &  9.4$^{+5.3}_{-3.7}$ & 330.0/444\\ 
2021-09-21 & 26134 & 1.16$^{+0.32}_{-0.32}$ & 6.5$_{-1.8}^{+2.4}$   &  5.4$^{+3.9}_{-2.1}$ & 298.8/444\\
2021-09-23 & 24300 & 1.25$^{+0.28}_{-0.28}$ & 8.3$_{-2.1}^{+2.5}$   &  6.4$^{+3.3}_{-2.3}$ & 289.8/444\\
2021-09-24 & 24303 & 1.69$^{+0.28}_{-0.28}$ & 13.2$_{-3.1}^{+3.6}$   & 6.9$^{+2.6}_{-2.3}$ & 292.2/444\\
\hline
\hline
\end{tabular}
\end{center}
}
{\footnotesize Notes: Galactic absorption ({\tt phabs} model) of N$_H= 1.82\times 10^{20}$cm$^{-2}$ was included in all model fitting with a power law model.
$^a$ \chandra\ obsid;
$^b$ Norm - normalization of a power law is defined at 1\,keV in 10$^{-5}$ photons s$^{-1}$cm$^{-2}$ keV$^{-1}$
$^c$ Flux in the 0.5-7~keV energy range in 10$^{-14}$erg~s$^{-1}$cm$^{-2}$; uncertainties are 90\%. }
\end{table}

\section{X-ray Properties of the Second X-ray Source}
\label{sec:Second-X-Appendix}


\begin{figure}
\includegraphics[width=0.45\columnwidth{}]{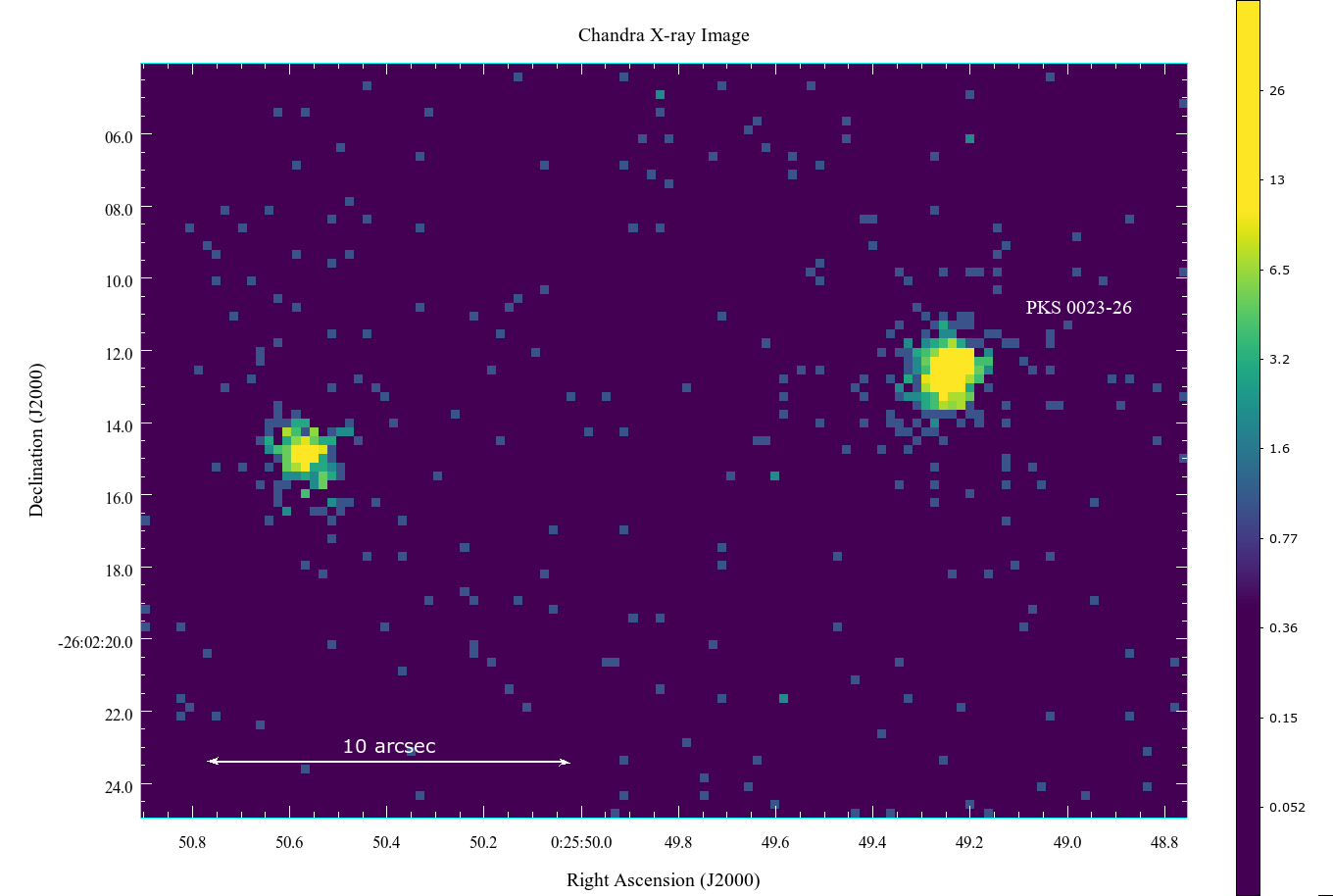} 
\includegraphics[width=0.4\columnwidth{}]{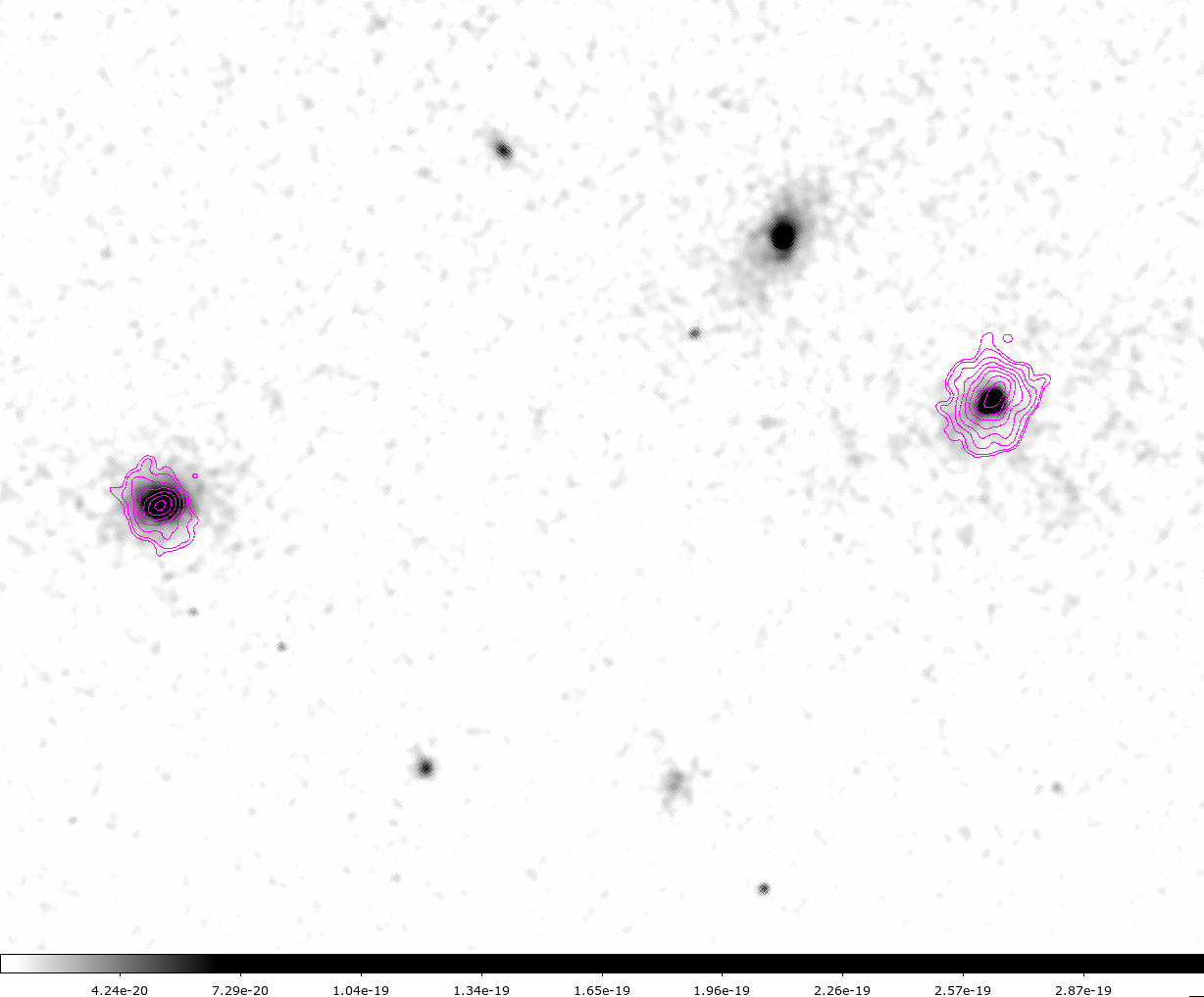}
\caption{Left: \chandra\ ACIS-S image in the 0.3-7 keV energy range binned to half of the ACIS pixel size of 0.\arcsec246. Two X-ray sources are seprated by about 18\arcsec. The scale is log with the color bars indicating counts per pixel. Right: Optical image from Gemini overplotted with the X-ray contours. The X-rays are associated with \pks\, and the second source to the East of \pks}
\label{fig:field}
\end{figure}

\begin{figure}
   \centering
\includegraphics[width=0.5\columnwidth{}]{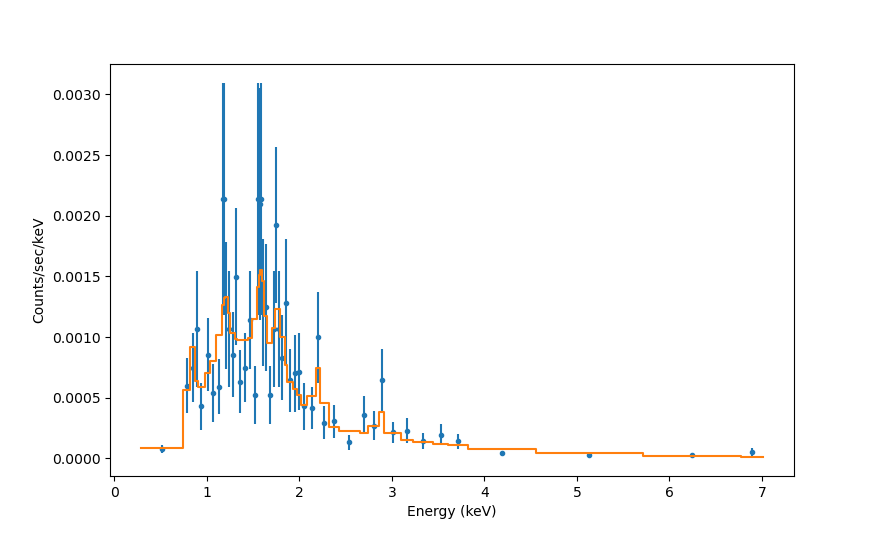}
\caption{X-ray spectrum (blue points) of the second source overplotted with the best fit model (solid orange line). The spectrum was grouped by 5 counts per bin for the visualization.
}
    \label{fig:spectrum-2s}
    \end{figure}

Figure~\ref{fig:field} shows an optical image from Gemini overplotted with the X-ray contours from the \chandra\ observation. In addition to the \pks\ X-ray source we also detect a second X-ray source located about 18\arcsec\, to the East. The source is present in the Gaia DR2 \citep{Gaia2018} optical catalog ID = 2323489174407213824 and the average G magnitude of 
20.6525$\pm$0.0211. 
This Gaia source is fainter in than \pks\ in X-rays and it was not separated in the XMM-Newton analysis presented in \cite{Mingo14} who used 30\arcsec\ extraction region for \pks\ analysis. We used {\tt{specextract}} tool to extract the X-ray spectra from each {\tt obsid} assuming the circular region centered on this source with the radius of 1.5$\arcsec$. These spectra were combined with {\tt{combined\_spectra}} to generate one spectrum that we fit using \sherpa. The total number of source net counts in the combined spectrum was 277.2$\pm16.8$ (background counts = 3.8$\pm1.9$). We fit the spectrum with a power law model and because the residuals were quite large we added several Gaussian lines to the model to obtain the best fit. Figure~\ref{fig:spectrum-2s} shows the best-fit model overplotted on the X-ray spectrum and
Table~\ref{tab:src2} presents the model parameters. We list the lines at the observed energy. The source does not have any redshift information in the literature and because several galaxies in the field have the redshift consistent with \pks\ we also assumed redshift of 0.322 for the emission lines.
we note that the power law continuum with measured photon index of $2.33^{+0.19}_{-0.28}$ is steeper than the X-ray photon index of \pks. We measure the soft-band X-ray flux of $f^{\rm \, cont}_{(0.5-2 \rm keV)} = 1.50^{+0.16}_{-0.18}\times 10^{-14} \, \rm erg~cm^{-2}~s^{-1}$ for the continuum power law only and $ f^{\rm \, full}_{(0.5-2 \rm keV)} =  1.81^{+0.21}_{-0.22}\times 10^{-14}\, \rm erg~cm^{-2}~s^{-1}$ for the continuum plus emission line model. The hard band flux for the power law continuum is equal to $f^{\rm \, cont}_{(2 -10 \rm keV)} = 1.04^{+0.24}_{-0.17}\times 10^{-14}\, \rm erg~cm^{-2}~s^{-1}$.

\begin{table}
	\caption{Properties of the Emission Lines in the Second X-ray Source.}
	\scriptsize
	\label{tab:src2}
	\begin{center}
	\begin{tabular}{ l c c c c c c}
		\hline \hline
 \\
 & E$_{\rm obs}$(keV)  & EW(keV) &  ID(obs) & E$_{\rm rest}$(keV) & ID(rest) \\
        \\
\hline
\\
   line 1           &    $0.82\pm0.05$ & $0.198_{-0.191}^{+0.101}$ & Fe XVII &  1.084 & Fe XXII \\
   line 2           &    $1.58\pm{0.03}$ & $0.124_{-0.027}^{+0.060}$ & Mg XI & 2.088 &  Si XIV \\
   line 3           &    $1.19_{-0.08}^{+0.78}$ &  $0.058_{-0.042}^{+0.028}$ & Fe XXIV & 1.57 & Mg XI \\
   line 4           &    $2.89_{---}^{+0.09}$ & $0.120_{-0.00}^{+0.130}$ & S XV & 3.82 & Ca XIX\\
   line 5           &    $2.20_{-0.05}^{0.04}$ & $0.241_{-0.078}^{+0.068}$ & Si XIII & 2.91  & S XV\\
   line 6           &    $1.76_{-0.05}^{0.12}$ & $0.075_{-0.014}^{+0.098}$ & Mg XII & 2.33 &  Si XIV \\
   \\
\hline
PowerLaw & $\Gamma=$   $2.33_{-0.28}^{+0.19}$ & Norm = $6.63_{-2.65}^{+0.61}$ \\
\hline
\\
		\end{tabular}
        
		Notes: Galactic absorption ({\tt phabs} model) of N$_H= 1.82\times 10^{20}$cm$^{-2}$ frozen in the fit; Redshift $z = 0.322$. The normalization of a power law model is defined at 1 keV and given in the units of $10^{-6}$ photons~cm$^{-2}$~s$^{-1}$~keV$^{-1}$. Equivalent width (EW)  in keV units and 1$\sigma$ errors; power law parameters are given in the last row with 90\% uncertainties.   
      \end{center}
\end{table}

\end{appendix}

\clearpage
\bibliography{all_refs}{}
\bibliographystyle{aasjournal}

\end{document}